\documentclass{article}
\usepackage[utf8]{inputenc}
\usepackage[margin=1in]{geometry}
\makeatletter
\providecommand{\@LN}[2]{}
\makeatother
\usepackage{times}
\usepackage{bm}
\usepackage{hyperref}
\hypersetup{
    colorlinks,
    linkcolor={red},
    citecolor={blue},
    urlcolor={blue!80!black}
}
\usepackage{url}
\usepackage{booktabs}
\usepackage{amsfonts,amssymb,amsmath}
\usepackage{nicefrac}
\usepackage{microtype}
\usepackage{graphicx}
\usepackage{bbm}
\usepackage{natbib}
\usepackage{color}
\usepackage{multirow}
\usepackage{caption}
\usepackage{subcaption}
\usepackage{authblk}
\usepackage{amsthm}
\usepackage{mathrsfs}
\usepackage{float}
\usepackage[subscriptcorrection]{newtxmath}
\usepackage[plain,noend]{algorithm2e}

\makeatletter
\renewcommand{\algocf@captiontext}[2]{#1\algocf@typo. \AlCapFnt{}#2}

\def\@algocf@capt@plain{top}
\renewcommand{\algocf@makecaption}[2]{%
  \addtolength{\hsize}{\algomargin}%
  \sbox\@tempboxa{\algocf@captiontext{#1}{#2}}%
  \ifdim\wd\@tempboxa >\hsize%
    \hskip .5\algomargin%
    \parbox[t]{\hsize}{\algocf@captiontext{#1}{#2}}%
  \else%
    \global\@minipagefalse%
    \hbox to\hsize{\box\@tempboxa}%
  \fi%
  \addtolength{\hsize}{-\algomargin}%
}
\makeatother

\newtheorem{theorem}{Theorem}

\newtheorem{lemma}{Lemma}

\graphicspath{ {./} }

\def\hat{\widehat}
\def\tilde{\widetilde}

\def\bSigma{\Sigma}

\def\bZ{{Z}}

\def\mux{\mu_{{}_W}}
\def\muxh{\hat\mu_{{}_W}}

\def\bbeta{{\beta}}
\def\bu{{u}}
\def\C{{\mathbb C}}
\def\V{{\mathbb V}}

\begin{document}

\title{\textbf{Shape-Preserving Covariate Adjustment via Empirical Likelihood in Randomized Experiments}}

\author{Zhilan Lou$^{1}$, Jun Shao$^{2}$, Yuhan Qian$^{3}$,  
Tuo Wang$^{4}$, Yanyao Yi$^{4}$, Yu Du$^{4}$, and  Ting Ye$^{3}$}
\affil{$^{1}$School of Data Sciences, Zhejiang University of Finance and Economics, Hangzhou, Zhejiang 310018, China.\\
$^{2}$Department of Statistics, University of Wisconsin, Madison, WI 53706, U.S.A.\\
$^3$Department of Biostatistics, University of Washington, Seattle, WA 98195, U.S.A.\\
$^{4}$Global Statistical Science, Eli Lilly and Company, Indianapolis, IN 46285, U.S.A.}

\maketitle

\begin{abstract}
Covariate adjustment improves estimation efficiency in randomized experiments, but standard calibration and augmentation methods, when applied to distribution or survival functions, do not preserve monotonicity---a fundamental property of the estimand. We propose using empirical likelihood with covariate-balancing constraints to construct a covariate-adjusted empirical measure for each treatment arm. Estimators of a broad class of distributional functionals, including cumulative distribution functions, survival functions, quantiles, and restricted mean survival times, are then derived as plug-in functionals of this measure, automatically inheriting proper shape constraints. We establish asymptotic normality with an explicit, guaranteed efficiency gain over unadjusted estimators.  The asymptotic distributions are invariant to the randomization scheme, providing a unified inference procedure under simple randomization and all commonly used covariate-adaptive designs satisfying a mild balancing condition.
This unified construction, adjusting the empirical measure once and deriving all estimators from it, offers a principled reconciliation of covariate adjustment with shape preservation. Simulations and an application to the SURPASS-4 trial confirm the theoretical gains.

\end{abstract}

\medskip
\noindent\textbf{Keywords:} 
Covariate-adaptive randomization, guaranteed efficiency gain, invariant asymptotic distribution,
survival function estimation.

\section{Introduction}
\label{sec: intro}

In randomized experiments comparing $J \geq 2$ treatments, interest often lies in the population distribution function $F_j(y) = P(Y_j \leq y)$ of an outcome  or a lifetime $Y_j$, or functionals thereof such as quantiles, interquartile ranges, rank-sum means, and restricted mean survival time. Baseline covariates, collected prior to treatment assignment, can improve estimation efficiency without relying on  models---an approach increasingly advocated by regulatory agencies \citep{ICHE9,ema:2015aa,fda:2019aa}. Existing model-free covariate adjustment methods, based on calibration or augmentation, treat $F_j(y)$ as a pointwise mean for each $y$ and adjust accordingly \citep{rao1990, zhang2008improving, zhang2015robust, Wang:2019aa, Liu:2020aa,%Ye:2020survival, 
cohen2021noharm, Wang:2021wg, ye2021better, ye2024logrank, Bannick}. This pointwise approach does not respect the global constraint that $F_j$ must be a distribution function: the resulting estimators can violate monotonicity  or be outside the unit interval, creating difficulties for interpretation and for downstream inference on quantiles and other distribution functionals.

 We propose a different approach: incorporating covariate adjustment directly as constraints in an empirical likelihood framework \citep{owen1988, qinlawless1994}, a well-established tool for nonparametric inference that naturally accommodates auxiliary information. Our contributions are fourfold. 
\begin{enumerate}
   \item We propose adjusting the empirical likelihood weights once, at the level of the empirical measure, rather than separately for each estimand. Specifically, under each treatment arm $j$, we compute probability weights by maximizing the empirical likelihood subject to the constraint that the weighted covariate mean matches the population mean. These adjusted weights are then plugged into a broad class of distributional functionals, including the weighted empirical distribution function, the weighted Kaplan-Meier estimator, quantile estimators, and treatment effect measures, all of which automatically inherit proper shape constraints. As a result, the maximum empirical likelihood estimator (MELE) of $F_j$ is always a valid distribution function, its survival function counterpart $S_j = 1-F_j$ is always a valid survival function, and the MELE of a quantile is always a proper quantile. This resolves the monotonicity violations that plague augmentation-based approaches \citep{rao1990, zhang2015robust}; we demonstrate empirically that augmented estimators violate monotonicity in {up to 100\%} of simulation runs depending on the setting. \vspace{1mm}
%    \item We show that the maximum empirical likelihood estimator (MELE) of $F_j$ is always a distribution function and the MELE of $S_j$ is always a survival function, resolving the monotonicity issue that plagues the augmentation approach in \cite{rao1990} and \cite{zhang2015robust} (demonstrated empirically: augmented estimators violate monotonicity in 6--100\% of simulation runs depending on the setting). For each treatment arm $j$, we compute probability weights $\hat p_{ij}$ by maximizing the empirical likelihood subject to the constraint that the weighted covariate mean matches the population mean. These weights, not estimands, are adjusted and determined once and then plugged into a broad class of distributional functionals---the weighted empirical distribution function, the weighted Kaplan-Meier estimator, quantile estimators, treatment effect measures---all of which automatically inherit proper shape constraints. \vspace{2mm}
    \item We establish that the asymptotic distributions of all proposed estimators are invariant to the randomization scheme: the same limiting distributions hold under simple randomization and under all covariate-adaptive designs satisfying a mild balancing condition~(D), including Pocock-Simon minimization, whose asymptotic properties remain incompletely understood. To handle the dependence structure induced by covariate-adaptive treatment assignment, we develop a dedicated asymptotic theory; both this theory and the resulting invariance property are, to our knowledge, new in the empirical likelihood literature.
    \vspace{1mm}
   % Among the nearest methodological competitors, the calibration-weighted estimators of \cite{rao1990} and the augmented estimators of \cite{zhang2015robust} share the goal of covariate-adjusted distribution and survival function estimation, but neither produces estimators with guaranteed monotonicity. 
   \item We prove a guaranteed efficiency gain: the asymptotic variance of the MELE is never larger than that of the unadjusted estimator, with an explicit expression for the gain in terms of covariate-outcome associations. The entropy balancing weights of \cite{zhao2017entropy} solve a dual problem that is asymptotically equivalent to ours for mean estimation, but have not been extended to function-valued estimands or to covariate-adaptive randomization. \cite{Wang:2021wg} establishes that the Kaplan-Meier estimator under stratified permuted block randomization is more efficient than under simple randomization, but does not incorporate covariate adjustment for further efficiency gain at the estimation stage.
     \vspace{1mm}
      \item  We establish a formal equivalence: the MELE, the augmented estimator based on outcome regression, and the entropy balancing estimator all share the same influence function for any square-integrable functional of $F_j$,  extending the scalar-estimand result of \cite{zhao2017entropy} to function-valued estimands under covariate-adaptive randomization. This equivalence clarifies that the three approaches differ not in asymptotic efficiency but in shape preservation: only the empirical likelihood and entropy balancing approaches guarantee valid distribution and survival function estimates. The empirical likelihood construction thus provides a unified framework that reconciles shape preservation, estimation-stage covariate adjustment, and design-invariant inference.
\end{enumerate}

%Our setting differs in two key respects. First, the estimands are function-valued (distributions and survival functions), not scalar means or estimating equation solutions, so that shape preservation---not available in the standard empirical likelihood framework for means---becomes the central advantage.

Section~2 describes covariate-adaptive randomization and the choice of adjustment covariates. Section~3 introduces the empirical likelihood weighting construction and derives covariate-adjusted estimators for distribution functions, survival functions, and other functionals. Section~4 develops the asymptotic theory, and Section~5 addresses variance estimation. Sections~6 and 7 present simulation studies and a real data application, respectively. All technical proofs are provided in the Supplementary Material.

\section{Covariate-Adaptive Randomization and Covariate Adjustment}

Consider a randomized experiment comparing $J \geq 2$ treatments. Let $(Y_1, \ldots, Y_J, X)$ denote the potential outcomes \citep{Neyman:1923a, Rubin:1974} and baseline covariate vector for a generic unit, and let $(Y_{i1}, \ldots, Y_{iJ}, X_i)$, $i = 1, \ldots, n$, be an independent and identically distributed sample from $(Y_1, \ldots, Y_J, X)$, where $Y_{ij}$ is the potential outcome of unit $i$ under treatment $j$ and $X_i$ is its baseline covariate vector. Each unit receives exactly one treatment, indicated by $A_i \in \{1, \ldots, J\}$, and under the consistency assumption the observed outcome is $Y_i = Y_{iA_i}$. Since the distribution of $X$ is unaffected by treatment assignment, covariate observations from all treatment groups, not just group $j$, can be leveraged to improve efficiency in estimating the outcome distribution $F_j(y) = P(Y_j \leq y)$.

Let $\pi_j$ denote the pre-specified target assignment proportion for treatment $j$, $j = 1, \ldots, J$, with $\pi_1 + \cdots + \pi_J = 1$. Under simple randomization, $A_1, \ldots, A_n$ are generated independently with $P(A_i = j) = \pi_j$, independent of the potential outcomes and $X$. While simple randomization ensures asymptotic validity of many statistical procedures, it does not exploit covariate information and may yield realized treatment proportions that deviate substantially from the targets $\pi_j$ within levels of important prognostic factors.

To address this, covariate-adaptive randomization has become standard practice, as discussed in Section~1; it adjusts for covariates at the treatment assignment stage rather than the estimation stage. The two most widely used schemes are stratified permuted block randomization \citep{Zelen:1974aa} and Pocock-Simon minimization \citep{Taves:1974aa, Pocock:1975aa}; reviews of these and other schemes are provided in \cite{Schulz:2018aa} and \cite{Shao:2021aa}. All commonly used covariate-adaptive randomization procedures incorporate a discrete baseline covariate $\bZ$ with finitely many levels as strata, and satisfy the following mild condition \citep{Baldi-Antognini:2015aa}.

\begin{description}
\item (D) Conditioned on $\bZ_1, \ldots, \bZ_n$, potential outcomes and covariates are independent of $A_1, \ldots, A_n$; the conditional probability $P(A_i = j \mid \bZ_1, \ldots, \bZ_n) = \pi_j$ for all $i$; and for every level $z$ of $\bZ$, the sequence $n_z^{1/2}\{n_{zj}/n_z-\pi_j\}$ is bounded in probability as $n\rightarrow\infty$, 
% $n_{zj}/n_z \to \pi_j$ in probability as $n \to \infty$, 
where $n_z$ is the number of units with $\bZ_i = z$ and $n_{zj}$ is the number of units with $\bZ_i = z$ and $A_i = j$.
\end{description}
Simple randomization is a special case of condition~(D), with $\bZ$ taken to be a constant.

Because covariate-adaptive randomization incorporates $\bZ$ into the generation of $A_1, \ldots, A_n$, it yields more efficient estimators than simple randomization. Further efficiency gains are possible at the estimation stage through adjustment for additional baseline covariates in $X$; for example, when components of $\bZ$ are discretized versions of continuous covariates, the original continuous covariates can be incorporated at the estimation stage to more fully exploit the available covariate information.

Let $W$ be a function of $X$ chosen for adjustment at the estimation stage. We require that $W$ have low dimension even if $X$ is high-dimensional, and that its covariance matrix $\bSigma = \mathrm{Var}(W)$ be finite and nonsingular to avoid collinearity. Under covariate-adaptive randomization, $A_1, \ldots, A_n$ are dependent on each other and on potential outcomes and covariates through $\bZ_1, \ldots, \bZ_n$. To handle this dependence in deriving the asymptotic properties of our estimators, we recommend including $\bZ$ in $W$; this is also a natural choice since $\bZ$ is the covariate used at the randomization stage. We say that $\bZ$ is included in $W$ if $W$ contains the indicator variables of all levels of $\bZ$.

%Let $\bX$ be the vector of covariates chosen for adjustment and $\bX_i$ be its value from unit $i$ in the sample.  

If a working linear model $E(Y_j \mid U) = \beta_j^\top U$ is available, where $U$ is a function of $X$ and $\beta_j$ is a parameter vector, then a natural choice is $W = (U^\top, \bZ
^\top)^\top$, combining $U$ with the randomization covariate $\bZ$. The working model need not be correctly specified. For outcomes where a generalized linear model is more appropriate, one may instead use the working model $E(Y_j \mid U) = g^{-1}(\beta_j^\top U)$, where $g$ is a known link function. Common examples include the logistic model, $g^{-1}(t) = e^t/(1+e^t)$, for binary outcomes, and the Poisson model, $g^{-1}(t) = e^t$, for non-negative integer-valued outcomes. In this case, we recommend $W = (h(U)^\top, \bZ^\top)^\top$,
where $h(U)$ is the $J$-dimensional vector whose $j$th component is $g^{-1}(\hat\beta_j^\top U)$ and $\hat\beta_j$ is estimated from the generalized linear model fit within treatment group $j$. Including $\bZ$ in $W$ is important for handling the dependence induced by covariate-adaptive randomization, even when $\bZ$ is already included in $U$. This choice of $W$ is motivated by the joint calibration approach of \cite{Bannick} and by the potential for substantial efficiency gains when the working model is approximately correct.

\section{Covariate Adjustment via Empirical Likelihood}

\subsection{The empirical likelihood weighting device}

The core of our approach is a single set of probability weights for each treatment arm, computed by maximizing an empirical likelihood subject to covariate balance constraints. These weights are the primary object; all subsequent estimators are derived as plug-in functionals of the resulting weighted empirical measure.

% For each $j = 1, \ldots, J$, we obtain weights $\hat{p}_{ij}$ by solving (fixing $\mu_W$)
For each fixed $\mux$, we define the profile empirical likelihood
 \begin{equation}\label{like}
 \ell (\mux) = \max_{\{p_{ij}\}} \sum_{j=1}^J \sum_{i: A_i=j} \log(p_{ij}), \quad \text{subject to} \quad p_{ij} \geq 0, \quad \sum_{i: A_i=j} p_{ij} = 1, \quad \sum_{i: A_i=j} p_{ij} W_i = \mux,
\end{equation}
where $p_{ij}$ is the probability mass at $(Y_{i}, W_i)$ conditional on $A_i = j$, $Y_{i}$ is the observed outcome and $W_i$ is the covariate vector for adjustment of unit $i$, and $\mux = E(W)$. The last constraint encodes covariate balance: since the distribution of $W$ is unaffected by treatment assignment, requiring the weighted covariate mean to equal the population mean $\mux$ across all arms simultaneously incorporates covariate information into the weights.
We estimate $\mux$ by the maximizer \(
\muxh =\arg\max_{\mux} \ell(\mux) 
\)
and define $\hat p_{ij}$ as the maximizer evaluated at $\muxh$.

By the method of Lagrange multipliers, the maximizers satisfy, for $j=1,...,J$ and $i=1,...,n$,  
\begin{equation}\label{phat}
\hat p_{ij} = \frac{I(A_i=j)}{n_j + \hat \lambda_j^\top (W_i  -  \muxh )}, \qquad
\sum_{j=1}^{J} \hat\lambda_j = 0 , \qquad  \sum_{i=1}^n  \frac{ I(A_i=j)(W_i - \muxh  )}{n_j+ \hat\lambda_j^\top (W_i - \muxh    )} = 0 ,
\end{equation}
where $I(B)$ is the indicator of event $B$, 
$n_j$ is the number of units in treatment group $j$,
and the $\hat\lambda_j$'s are Lagrange multipliers.

The empirical likelihood weights~(\ref{phat}) exist and are unique with probability tending to 1 as $n \to \infty$. This follows from the standard empirical likelihood existence result \citep{owen1988, qinlawless1994}: the solution is well-defined whenever $\mux$ lies in the interior of the convex hull of $\{W_i : A_i = j\}$, and under condition~(D) with $\mathrm{Var}(W)$ nonsingular, this event occurs with probability tending to 1 by the law of large numbers. In finite samples, particularly with small treatment arms or moderately high-dimensional $W$, the convex hull condition may fail; standard remedies are available in such cases \citep{chen2008adjusted}.

%The efficiency gain $\pi_j^{-1}(1-\pi_j) {\mathbb C}_j(y)^\top \Sigma^{-1} {\mathbb C}_j(y) \geq 0$ by construction as a quadratic form, and is strictly positive whenever ${\rm Cov}\{W, I(Y_j \leq y)\} \neq 0$.

\subsection{Distribution function estimation: uncensored outcomes}

Without covariate adjustment, the standard estimator of $F_j(y) = P(Y_j \leq y)$ is the empirical distribution function
\begin{equation}\label{fhatu}
\tilde{F}_j(y) = \frac{1}{n_j} \sum_{i: A_i=j} I(Y_i \leq y), \qquad -\infty < y < \infty, \quad j = 1, \ldots, J.
\end{equation}
Replacing the uniform weights $1/n_j$ with the empirical likelihood weights $\hat{p}_{ij}$ from \eqref{phat} yields the covariate-adjusted MELE of $F_j(y)$:
\begin{equation}\label{fhat}
\hat{F}_j(y) = \sum_{i: A_i=j} \hat{p}_{ij} I(Y_i \leq y), \qquad -\infty < y < \infty, \quad j = 1, \ldots, J.
\end{equation}
Since $\hat{p}_{ij} \geq 0$ and $\sum_{i: A_i=j} \hat{p}_{ij} = 1$ by the 
empirical likelihood constraints, $\hat{F}_j$ is guaranteed to be a valid 
distribution function: nondecreasing with $\hat{F}_j(-\infty) = 0$ and 
$\hat{F}_j(\infty) = 1$.

\cite{rao1990} proposed the augmented (calibration) estimator
\begin{equation}\label{fhata}
\hat{F}^{\rm A}_j(y) = \tilde{F}_j(y) + \frac{1}{n} \sum_{i=1}^n I(\hat{\beta}_j^\top W_i \leq y) -
\frac{1}{n_j} \sum_{i: A_i=j} I(\hat{\beta}_j^\top W_i \leq y),
\end{equation}
where $\hat{\beta}_j = \hat{\Sigma}_j^{-1} \sum_{i: A_i=j} (W_i - \bar{W}_j) Y_i / n_j$, and $\bar{W}_j$ and $\hat{\Sigma}_j$ are the sample mean and covariance matrix of $\{W_i : A_i = j\}$, respectively.
$\hat{F}^{\rm A}_j(y)$ is not guaranteed to be monotone in $y$: the last two terms of \eqref{fhata} jointly drop by $n_j^{-1}-n^{-1}$ at each $y=\hat\beta_j^\top W_i$ with $A_i=j$, so with a continuous covariate $\hat F^{\rm A}_j$ is non-monotone in every sample. In our simulations reported in the Supplementary Material, $\hat{F}^{\rm A}_j(y)$ is non-monotone in 100\% of 2,000 runs.

Estimators of functionals of $F_j$ are obtained by applying the same functional to 
$\hat{F}_j$ in \eqref{fhat}. For example, the MELE of the population mean $E(Y_j) = 
\int y\, dF_j(y)$ is $\hat{\mu}_j = \int y\, d\hat{F}_j(y) = \sum_{i: A_i=j} \hat{p}_{ij} Y_i$, 
and the MELE of the quantile $F_j^{-1}(p)$ at a fixed $p \in (0,1)$ is $\hat{F}_j^{-1}(p)$, 
where $G^{-1}(p) = \inf\{y : G(y) \geq p\}$ for any nondecreasing $G(y) \in [0,1]$. The 
corresponding unadjusted estimators are the same functionals of $\tilde{F}_j$: the sample 
mean $\int y\, d\tilde{F}_j(y)$ and sample quantile $\tilde{F}_j^{-1}(p)$. Since $\hat{F}_j$ 
is a valid distribution function by construction, $\hat{F}_j^{-1}(p)$ is always a proper 
quantile. More generally, any Hadamard-differentiable functional of $\hat{F}_j$ or 
$\tilde{F}_j$ inherits their asymptotic properties via the functional delta method.

 To compare two treatments $j \neq k$, the MELE  of the population mean difference $E(Y_j)-E(Y_k)$ is 
 $\hat{\mu}_j - \hat{\mu}_k$, and the MELE  of the population quantile difference $F_j^{-1}(p)-F_k^{-1}(p)$  is $\hat F_j^{ -1}(p)-\hat F_k^{ -1}(p)$. For continuous $F_j$ and $F_k$, another treatment effect measure is $ \theta_{jk} = \int \! \! \int I(y_j \leq y_k)d F_j(y_j) d F_k(y_k)$, which is the population quantity targeted by the Wilcoxon–Mann–Whitney test.  Its MELE is
$ \hat \theta_{jk} \!=\!  \int \! \! \int I(y_j \leq y_k)d \hat F_j(y_j) d \hat F_k(y_k) \!=\!  \sum_{i: A_i =j} \sum_{i': A_{i'} =k} \hat p_{ij} \hat p_{i'k} I(Y_{ij} \leq Y_{i'k})$ and the corresponding 
unadjusted estimator is the  Wilcoxon two sample statistic
$\tilde \theta_{jk} = n_j^{-1}n_k^{-1}\! \sum_{i: A_i =j} \! \sum_{i': A_{i'} =k}  I(Y_{ij} \!\leq \!Y_{i'k})$.

\subsection{Survival function estimation: censored outcomes}

We now consider the setting where the outcome $Y_j \geq 0$ is a failure time subject 
to right-censoring by $C_j > 0$ under treatment $j = 1, \ldots, J$. We assume that $Y_j$ and $C_j$ are independent given $A = j$, the standard assumption 
underlying the Kaplan-Meier estimator. For unit $i$ assigned to arm $A_i = j$, let $Y_i = Y_{ij}$ and $C_i = C_{ij}$ denote the failure and censoring times under the assigned treatment; 
we observe $\min(Y_i, C_i)$ and the event indicator 
$I(Y_i \leq C_i)$. 
The quantity of interest is the survival function 
$S_j(t) = P(Y_j \geq t)$ for $t \geq 0$.

For unit $i$, define the counting and at-risk processes $N_{ij}(t) = I(A_i = j,\, Y_i \leq C_i,\, Y_i \leq t)$ and 
$Y_{ij}(t) = I(A_i = j,\, Y_i \geq t,\, C_i \geq t)$, with population-level analogues $N_j(t)=I(A=j,\;Y_j\le C_j,\;Y_j\le t)$ and $Y_j(t)=I(A=j,\;Y_j\ge t,\;C_j\ge t)$.
Without covariate
adjustment, the Kaplan-Meier estimator \citep{kaplan1958nonparametric} of $S_j(t)$ is
\begin{equation}\label{KM}
\tilde{S}_j(t) = \prod_{s \leq t} \Big\{ 1 - \frac{\sum_{i: A_i = j} dN_{ij}(s)}{\sum_{i:  A_i = j} Y_{ij}(s)} \Big\}. 
\end{equation}
 In the absence of censoring ($C_j \equiv \infty$), 
$\tilde{S}_j = 1 - \tilde{F}_j$, where $\tilde{F}_j$ is defined in \eqref{fhatu}.

%\citep{kalbfleisch2011statistical}.
%$\tilde S_j (t)$ for each $t$  is consistent and asymptotically normal

%However, the asymptotic distribution of $\hat S_j^{\rm KM} (t)$ varies with covariate-adaptive randomization of generating $A_1,...,A_n$  \citep{Wang:2021wg}. 
%Pocock-Simon's minimization
%An efficiency gain by adjusting covariates can be achieved either  in the treatment assignment  randomization process or in the estimation stage.

Using the same empirical likelihood weights $\hat p_{ij}$ from~(\ref{phat}), our proposed MELE of $S_j(t)$ is the weighted Kaplan-Meier estimator 
\begin{equation}
\hat S_j(t) = \prod_{s \leq t} \Big\{ 1- \frac{\sum_{i: A_i = j} \hat p_{ij}  dN_{ij}(s)}{\sum_{i: A_i = j} \hat p_{ij}  Y_{ij}(s) } \Big\} , \label{AKM}
\end{equation}
where $d\hat \Lambda_j(t)= \frac{\sum_{i: A_i = j} \hat p_{ij}  dN_{ij}(t)}{\sum_{i: A_i = j} \hat p_{ij}  Y_{ij}(t) } $ and $\hat\Lambda_j(t)=\int_0^td\hat\Lambda_j(s)$ is the corresponding weighted Nelson-Aalen estimator of the cumulative hazard. It is clear that $\hat{S}_j(t)$ is nonincreasing with $\hat{S}_j(0) = 1$. 
It reduces to the standard Kaplan-Meier estimator when $\hat{\lambda}_j = 0$ 
(no covariate adjustment), and to $1 - \hat{F}_j$ in the absence of censoring.

\begin{figure}[ht]
    \centering
    \includegraphics[scale=0.6]{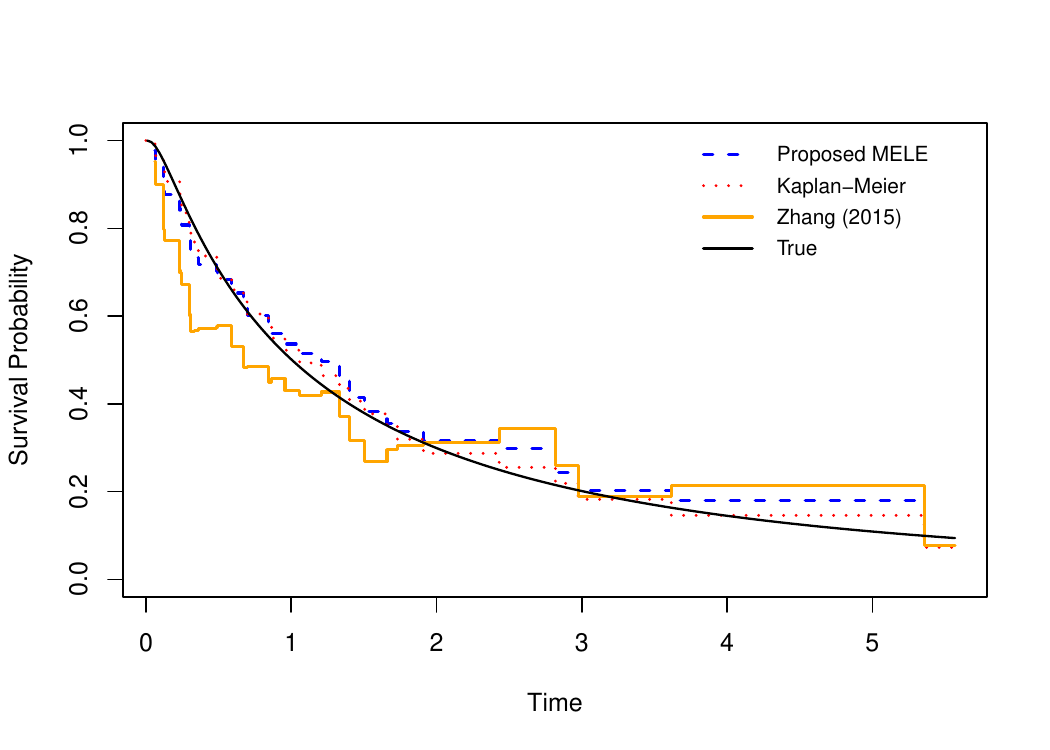}
    \caption{Three estimated survival curves under simple randomization.}
    \label{fig:case1-simple-nonmonotone}
\end{figure}

 Under simple randomization, \cite{zhang2015robust} estimated the cumulative hazard increment $d\Lambda_j(t)$ by 
 $d\hat\Lambda_j^{\rm AGEE}(t)$ obtained from the augmented generalized estimating equation (AGEE)
\begin{equation}
\sum_{i=1}^n \left[ \{dN_{ij}(t) - Y_{ij}(t)\, d\Lambda_j(t)\} - \{I(A_i = j) - \pi_j\}(W_i - \mu_W)^\top \hat\beta_j(t)\, dt \right] = 0,
\end{equation}
where $
\hat\beta_j(t)\, dt = n_j^{-1} \hat\Sigma_j^{-1} \sum_{i: A_i = j} (W_i - \bar W_j) \left\{ dN_{ij}(t) - Y_{ij}(t)\, d\tilde \Lambda_j(t) \right\}$, $\bar W_j$ and $\hat\Sigma_j$ are given in \eqref{fhata}, and $\tilde \Lambda_j(t)$ is the unadjusted Nelson–Aalen
estimator. The AGEE survival estimator is then the product integral $\prod_{s \le t}\{1 - d\hat\Lambda_j^{\rm AGEE}(s)\}$.
However,  the AGEE curve is not necessarily a monotone function, like the estimator $\hat F_j^{\rm A} $ in (\ref{fhata}). 
In the simulation results in Section 6, the AGEE curve is not monotone in about 3\% of 2,000 runs.  
One example under simple randomization is given in Figure \ref{fig:case1-simple-nonmonotone}, where the AGEE curve  increases at about 10 time points. Although monotonicity could in practice be restored by post-processing the AGEE estimator via isotonic regression or the rearrangement operator of \citet{chernozhukov2010}, such modifications alter the pointwise asymptotic distribution and require a separate, more delicate inferential theory. By contrast, the empirical likelihood approach produces a proper empirical measure from the outset, so all downstream functionals inherit both shape constraints and a unified asymptotic theory.

For a time-to-event outcome with censoring, the treatment effect comparing treatments $j$ and $k$ is sometimes assessed by the difference  
\[ 
\Delta_{jk}  = \! \int_0^{\tau} \!\! \{ S_j(t) -
 S_k(t) \} dt  ,
\]
%\[ \D = {\rm AUC}_j- {\rm AUC}_k, \quad \ \  {\rm AUC}_l  = \! \int_0^{\tau}\!\! S_l(t) dt = \! \int_0^{\tau} \! (\tau - t) S_l(t) dR_l(t) , \quad l=j,k, \]
where $\int_0^{\tau} \!  S_l(t)dt $ is the restricted mean survival time for treatment $l$ \citep{claggett2022,sun2025} 
%(RMST) under   %also called  in \cite{},   $dR_l(t) = E \{ dN_l(t) \mid \min(Y_l , C_l ) \geq t \}$, 
and $\tau$ is a fixed point such that $P \{ \min (Y_l, C_l ) \geq \tau \mid A=l \} >0 $, $l= j,k$. 
The covariate adjusted MELE $\hat \Delta_{jk}$ and the unadjusted estimator   $\tilde \Delta_{jk}$  are  obtained by substituting 
$S_l(t)$  in $\Delta_{jk}$ by $\hat S_l(t) $ and $\tilde S_l(t)$, respectively, $l=j,k$. 

\section{Asymptotic Theory}
\subsection{Distribution estimators} 
\label{sec: asymptotic}

We derive the asymptotic distributions for the covariate adjusted 
MELE $\hat F_j$ in (\ref{fhat}) and 
$\hat S_j$ in (\ref{AKM}), under the same condition for unadjusted estimators plus the minor condition (D) satisfied by all commonly used covariate-adaptive randomization schemes.

\begin{lemma}\label{lemma1}
	Assume (D). For $\muxh$ and $\hat\lambda_j$, $j=1,...,J$, defined in (\ref{phat}), if  $E\|W\|^3<\infty$,
	\[ \muxh = \bar W + n^{-1/2} o_p(1) \quad \mbox{and} \quad 	
	 n^{-1}\hat{\lambda}_j = n^{-1}\sum_{i=1}^{n} \left\{I(A_i=j) - \pi_j\right\} \Sigma^{-1} (W_{i} - \mux ) + n^{-1/2} o_p(1), \]
	where 	$o_p(1)$ denotes a quantity converging to 0 in probability as $n\to \infty$, and $\bar W$ is the sample mean of all $W_i$'s.
\end{lemma}

Lemma~\ref{lemma1} provides linearizations of $\muxh$ and $\hat\lambda_j$ that underlie all subsequent results. {Although the derivation builds on established empirical likelihood techniques, two aspects are non-standard. First, the constraints are tied to a common covariate mean that is estimated through a profile empirical likelihood and shared across treatment arms, so the convergence rates of the multipliers and of the fitted mean are obtained jointly rather than separately. Second, the analysis is conducted under covariate-adaptive randomization satisfying condition (D), under which treatment assignments are dependent.}

\begin{theorem}
Assume (D) with  $\bZ$ included in $W$ for adjustment in estimation and $E\|W\|^3<\infty$. \\
(i)  Result for $\hat F_j$ in (\ref{fhat}) without censoring. For any real $y$ and any $j$, as $n \to \infty$, 
$$%	\begin{equation}\label{r1}
		\sqrt{n} \{ \hat F_j (y) -F_j(y)\}  =  \frac{1}{\sqrt{n}}  \sum_{i=1}^n \phi_{ij} (y) + o_p(1) \  \stackrel{d}{\to} \ N \big(0, \, V_{F_j} (y) \big)
$$%	\end{equation}
	regardless of which randomization scheme is used,
	where
	$$
	\phi_{ij}(y)  = [ I(A_i=j)\{ I(Y_{ij} \leq y) - F_j(y)  \}  - \{ I(A_i=j) - \pi_j\} {\mathbb C}_{j}(y)^\top \Sigma^{-1} (W_i- \mux )] \big/ \pi_j,  
	$$
$\Sigma = {\rm Var}(W)$, ${\mathbb C}_{j}(y)$ is the  vector ${\rm Cov} \{W,  \, I(Y_j \leq y)\}$,  
	$\stackrel{d}{\to}$ is convergence in distribution, and %$N\big(0,V_{F_j}(y)\big)$ is the normal distribution with mean 0 and variance 
	$$V_{F_j} (y)= [ F_j(y) \{1 - F_j(y)\} - (1-\pi_{j}) {\mathbb C}_{j}(y)^\top \Sigma^{-1}   {\mathbb C}_{j}(y)] \big/ \pi_j. 
	$$
 (ii) Result for $ \hat S_j$ in (\ref{AKM}) with censoring. Under condition 
 \begin{description}
 	\item (C) $ \ C_A $ and $Y_A $ are independent conditioned on $ A$ and, 
 	for each $j$, $S_j(t)$ is continuous in $t$ and
 	there is a $\tau_j > 0$ such that $P\{ \min (Y_j,C_j)\geq \tau_j \mid A = j\} >0$, 
 \end{description}
for  any $t \in [0, \tau_j]$ and $j$, as $n \to \infty$,  
$$%\begin{equation}
\sqrt{n} \{  \hat S_j(t) - S_j(t) 
\}  =  \frac{1}{\sqrt{n}}  \sum_{i=1}^n \varphi_{ij} (t) + o_p(1) 
\ \stackrel{d}{\to} \ N\big(0, V_{S_j}(t)\big) , 
$$%\end{equation}
regardless of which randomization scheme is used,
where  
$$
\varphi_{ij}(t)  = - S_j(t) \! \! \int_0^t \! \frac{dM_{ij}(s)}{ E\{ Y_{j}(s)\} }  + \{ I(A_i=j)-\pi_j\}\Gamma_j(t)^\top \Sigma^{-1}
(W_i -\mux )\big/ \pi_j, $$
$dM_{ij} (t)  = dN_{ij} (t) - Y_{ij} (t) d \Lambda_j(t)$,
$\Gamma_j (t) =
{\rm Cov} \big(W, \, S_j(t)  \int_0^t  dM_{j}(s)/E\{ Y_{j}(s)\} \big)$, 
$dM_{j} (t)  = dN_j(t)  - Y_j(t) d\Lambda_j(t)$,  and
$$ V_{S_j}(t)  = 
{\rm Var} \bigg\{S_j(t) \!\! \int_0^t \! \frac{dM_{j}(s)}{E\{ Y_{j}(s)\}}  \bigg\} - (1-\pi_j) \Gamma_j(t)^\top  \Sigma^{-1} \Gamma_j(t)\big/\pi_j.$$
\end{theorem}
\vspace{2mm}

The proof (in the Supplementary Material) applies a special technique to handle the dependence in sequence $Y_{iA_i}, A_i, W_i$, $i=1,...,n$, due to  
covariate-adaptive randomization.  Note that condition (C) is also required for the unadjusted Kaplan-Meier estimator \citep{kalbfleisch2011statistical}. 

The following are elaborations of the results in Theorem 1. 
\begin{enumerate}
	\item 
	Theorem 1 shows not only the asymptotic validity of covariate adjusted $\hat F_j$  and $\hat S_j$ 
	but also the invariance of their asymptotic distributions, i.e., the same formula  holds for simple randomization or any  covariate-adaptive  randomization satisfying (D) including Pocock-Simon's minimization whose asymptotic property is still not well understood.
	The  invariant asymptotic distribution provides a unified formula for variance estimation and inference under all commonly used randomization,  desirable for  practitioners. 
	\item Under simple randomization, $\sqrt{n} \{  \tilde F_j(y) - F_j(y)\}  \stackrel{d}{\to} N\big(0,  \, F_j(y)\{ 1- F_j(y)\}/\pi_j\big) $. Thus, Theorem 1(i) shows that the MELE $\hat F_j(y)$  has a guaranteed efficiency gain over  the unadjusted $\tilde F_j(y)$ and the explicit  gain is $\pi_j^{-1}(1-\pi_j)   \C_{j}(y)^\top \Sigma^{-1}   \C_{j}(y)$.
	Similarly, under simple randomization, $\sqrt{n} \{  \tilde S_j(t) - S_j(t)\}  \!\stackrel{d}{\to} \!N\big( 0, {\rm Var} \{S_j(t) \!  \int_0^t \! dM_{j}(s)/E\{Y_{j}(s)\}\}   \!\big)$ \citep{kalbfleisch2011statistical} and Theorem 1(ii)  shows that the MELE $\hat S_j(t)$  has a guaranteed efficiency gain over  the unadjusted  Kaplan-Meier estimator $\tilde S_j(t)$ with explicit efficiency gain $\pi_j^{-1}(1-\pi_j) \Gamma_j(t)^\top  \Sigma^{-1} \Gamma_j(t)$.
	\item Consider the special case where $W=Z$. %, i.e., no covariate  other than those used in randomization is adjusted in estimation. 
    For  covariate-adaptive randomization satisfying  $n_z^{-1/2} (n_{zj} - \pi_j n_z) = o_p(1)$ for all $z$ with $n_{zj}$ and $n_z$  given in (D),  e.g., the stratified permuted block randomization, 
	covariate adjustment  in the estimation stage with $W=Z$ does not provide any additional efficiency gain over the  adjustment done in randomization, 
	 i.e., the  asymptotic distributions of $\hat F_j(y)$ and $ \tilde F_j(y)$   (or $\hat S_j(t)$ and $ \tilde S_j(t)$)  are the same when $W=Z$.
The result for Kaplan-Meier estimator $ \tilde S_j(t)$ under stratified permuted block randomization is actually obtained in   \cite{Wang:2021wg}. However, the same conclusion for $ \tilde S_j(t)$ does not hold under
Pocock-Simon's minimization, due to the complex asymptotic behavior of  $n_z^{-1/2} (n_{zj} - \pi_j n_z) $ under  minimization. 
%In fact, no result for $\tilde S_j(t)$ under  randomization scheme other than simple randomization or stratified permuted block randomization can be found in the literature. Also, there is no result in the literature for  survival function  estimation 
In contrast, results in Theorem 1 hold  for $\hat F_j(y)$ and $\hat S_j(t)$ under minimization when  covariate adjustment in estimation is also applied. 
\item If some levels of $Z$ are not in $W$, then results in Theorem 1 do not hold;  the asymptotic distribution of $\hat F_j(y)$ or $\hat S_j(t)$
	 varies with the particular randomization scheme and is not available under Pocock-Simon's minimization.
	Thus, we recommend including indicators for all levels of $Z$ in $W$.
\end{enumerate}

\subsection{Asymptotic equivalence among the empirical likelihood, augmentation, and entropy balancing}

The following theorem unifies three distinct covariate adjustment approaches,
the empirical likelihood, augmentation, and entropy balancing \citep{zhao2017entropy}, under one asymptotic result under covariate-adaptive randomization. 

\begin{theorem}[Asymptotic equivalence]\label{prop:equiv}
	Assume (D) with $\bZ$ included in $W$. Let $g(Y_j)$ be a function  with $E \{g(Y_j)^2\} \! < \! \infty$, $\theta_j \! = \! E\{g(Y_j)\}$,  $\C_j^g \! = \!{\rm Cov}\{W, g(Y_j)\}$, and
  $\hat \theta_j^{\star}$ be one of the following:\\
(a)  the proposed MELE $ \hat\theta_j = \int g(y) d \hat F_j(y) = \sum_{i: A_i = j} \hat p_{ij} \, g(Y_{ij})$; \\
	(b) the augmentation estimator $\hat \theta_j^{\rm \ AUG} = n_j^{-1}\sum_{i: A_i=j} g(Y_{ij}) - \hat \gamma_j^\top (\bar W_j - \bar W)$,
		where $\bar W_j = n_j^{-1}\sum_{i: A_i=j} W_i$, $\bar W = n^{-1}\sum_{i=1}^n W_i$, and $\hat \gamma_j = n_j^{-1} \hat \Sigma_j^{-1} \sum_{i: A_i=j} (W_i - \bar W_j) g(Y_{ij})$; \\
	(c)  {\color{black} the entropy balancing estimator $\hat \theta_j^{\rm \ EB} = \sum_{i: A_i = j} \hat p_{ij}^{\rm \ EB} \, g(Y_{ij})$, where, for each treatment arm \(j\), the entropy-balancing weights \(\{\hat p_{ij}^{\rm EB}:A_i=j\}\) solve $\max_{\{p_{ij}:A_i=j\}}-\sum_{i:A_i=j}p_{ij}\log p_{ij}$ subject to $p_{ij}\ge0$, $\sum_{i:A_i=j}p_{ij}=1$, and $\sum_{i:A_i=j}p_{ij}W_i=\bar W$.}\\
Then, {under the condition $E\|W\|^3<\infty$}, regardless of which covariate-adaptive randomization  scheme is used, 
$$
\sqrt{n}\big(\hat\theta_j^{\star} - \theta_j \big) = \frac{1}{\sqrt{n}} \sum_{i=1}^n \psi_{ij}^g + o_p(1) \, \stackrel{d}{\to} \, N\Big(0, \, [{\rm Var}\{g(Y_j)\} - (1-\pi_j)\C_j^{g\top} \Sigma^{-1} \C_j^g]/\pi_j\Big), 
$$
where 
$
\psi_{ij}^g = [I(A_i=j)\{g(Y_{ij}) - \theta_j\} - \{I(A_i=j) - \pi_j\} \C_j^{g\top} \Sigma^{-1}(W_i - \mux)]\big/\pi_j$.
\end{theorem} \vspace{2mm}

If $g(y) = I(y \leq t)$ in Theorem \ref{prop:equiv}, then we recover the asymptotic result in Theorem 1 for $ \hat F_j$, and obtain the asymptotic equivalence among $\hat F_j$, $\hat F_j^{\rm \ AUG} $, and $\hat F_j^{\rm \ EB} $. 
If $g(y) = y$, then all three estimators in Theorem \ref{prop:equiv} 
are asymptotically equivalent to the ANHECOVA estimator in \cite{ye2021better} for the mean $E(Y_j)$.

%Substituting into $\hat\theta_j^{\rm EL}$,
%\begin{align*}
%\hat \theta_j^{\rm EL} &= \frac{1}{n_j}\sum_{i:A_i=j} g(Y_{ij}) - \frac{1}{n_j}\sum_{i:A_i=j} g(Y_{ij}) \frac{\hat\lambda_j^\top(W_i - \muxh)}{n_j} + o_p(n^{-1/2}) \\
%&= \frac{1}{n_j}\sum_{i:A_i=j} g(Y_{ij}) - \widehat{{\rm Cov}}_j\{W,g(Y_j)\}^\top \frac{\hat\lambda_j}{n} + o_p(n^{-1/2}),
%\end{align*}
%where $\widehat{{\rm Cov}}_j$ denotes the within-arm sample covariance. By Lemma 1, $\hat\lambda_j/n = \Sigma^{-1}\bar U_j + o_p(n^{-1/2})$ with $\bar U_j = n^{-1}\sum_{i=1}^n\{I(A_i=j)-\pi_j\}(W_i - \mux)$. By the law of large numbers, $\widehat{{\rm Cov}}_j\{W,g(Y_j)\} \to C_j^g$ and $\bar U_j = (\bar W_j - \bar W)\pi_j + o_p(n^{-1/2})$. Thus,
%$$\hat\theta_j^{\rm EL} = \frac{1}{n_j}\sum_{i:A_i=j}g(Y_{ij}) - (C_j^g)^\top\Sigma^{-1}\bar U_j + o_p(n^{-1/2}).$$
%An identical expansion holds for $\hat\theta_j^{\rm aug}$ by direct algebra, since $\hat\gamma_j \to \Sigma^{-1}C_j^g$ and $\bar W_j - \bar W = \pi_j^{-1}\bar U_j + o_p(n^{-1/2})$.
%For entropy balancing, the dual of the constrained optimization has the same first-order conditions as empirical likelihood \citep{zhao2017entropy}, so the weight linearization is identical. The asymptotic distribution then follows from the central limit theorem for covariate-adaptive randomization \citep{ye2021better}. \hfill $\square$

Theorem~\ref{prop:equiv} shows that the three distinct approaches 
share the same influence function and are first-order asymptotically equivalent for any fixed function $g$. The critical distinction is in finite-sample behavior for function-valued estimands. When $g(y) = I(y \leq t)$ and one considers estimating $ F_j(t)$ as a function of $t$, the augmented estimator $\hat F_j^{\rm \, AUG}(t)$ applies the equivalence pointwise and does not enforce monotonicity across $t$-values, because the regression coefficient $\hat\gamma_j$ varies with $t$. Note that $\hat F_j^{\rm \, AUG}(t)$ is different from $\hat F_j^{\rm A}$ in \eqref{fhata} but both have the same issue.
The empirical likelihood approach, by contrast, uses the \emph{same} weights $\hat p_{ij}$ for all $t$, which is what ensures that $\hat F_j(t) = \sum_{i:A_i=j}\hat p_{ij}I(Y_{ij}\leq t)$ is automatically nondecreasing. This also holds for the entropy balancing approach.  Thus, among the three asymptotically equivalent methods, the empirical likelihood and entropy balancing preserve shape constraints while the augmentation does not.
The same reasoning extends to the censored setting: the AGEE of \cite{zhang2015robust} is asymptotically equivalent to our MELE $\hat S_j(t)$, but uses $t$-varying regression coefficients $\hat\beta_j(t)$ and does not produce a monotone survival curve estimator.

\subsection{Estimators of treatment effects}

We now consider the comparison of two treatments $j$ and $k$. 
For the mean treatment effect $E(Y_j)- E(Y_k)$, the asymptotic distribution of the MELE $\hat \mu_j - \hat \mu_k$ is a direct corollary of Theorem 1.  Thus, we consider the quantile treatment effect $F_j^{-1} (p) - F_k^{-1}(p)$ for a fixed known $ p \in (0,1)$, the rank-sum mean $\theta_{jk}$ defined in the end of Section 3.2, and the difference of restricted mean survival times $\Delta_{jk} $ defined in the end of Section 3.3. Let $\hat F_j^{-1}(p)-\hat F_k^{-1}(p)$, $\hat\theta_{jk}$, and $\hat \Delta_{jk}$ be  their respective MELEs. 

\begin{theorem}
	Assume (D) with  $\bZ$ included in $W$ and {the condition $E\|W\|^3<\infty$.} \\
  (i)  Assume that $F_l$ has a positive derivative $f_{l p}$ at $F^{-1}_l (p)$, $l=j,k$. For non-censored outcome, 
$$
\sqrt{n} \{ \hat F_{j}^{-1}(p)- \hat F_k^{-1} (p)- F^{-1}_j(p)+ F_k^{-1}(p)\} \  \stackrel{d}{\to} \ N\left(0, \
\frac{ p(1-p)}{\pi_j f_{jp}^2} +  \frac{p(1-p)}{\pi_k f_{kp}^2} - \xi_{jk} \right) 
$$
	regardless of which covariate-adaptive randomization scheme is used,
 where
 {$$ \xi_{jk}   =   \frac{(\pi_j \tilde\C_{k} + \pi_k \tilde\C_{j})^\top \Sigma^{-1}(\pi_j \tilde\C_{k} + \pi_k \tilde\C_{j}) }{\pi_j\pi_k (\pi_j+\pi_k)} + \frac{(1-\pi_j-\pi_k)(\tilde\C_{j} - \tilde\C_{k})^\top \Sigma^{-1}(\tilde\C_{j} - \tilde\C_{k})  }{(\pi_j+\pi_k)},$$
$\tilde \C_{l}=\C_{l}f_{lp}^{-1}$, and $\C_{l}= {\rm Cov} \{W,  \, I(Y_l \leq  F^{-1}_l(p))\}$, $l=j,k$.\\}
 % The result  also holds for censored outcome with $\C_{l}$ replaced by  ${\rm Cov} \big( W, (1-p) \int_0^{F_l^{-1}(p)} \! d M_l(s)/E\{ Y_l (s)\}  \big)$,  $l = j, k$. \\
(ii) Assume that $F_l$ is continuous, $l=j,k$. For non-censored outcome, 
\[ \sqrt{n} ( \hat \theta_{jk} - \theta_{jk} ) \  \stackrel{d}{\to} \ N\left(0, \
\frac{ {\rm Var }\{ F_k(Y_j)\}}{\pi_j } +  \frac{{\rm Var} \{F_j(Y_k)\} }{\pi_k } - \zeta_{jk}  \right)  \] 
	regardless of which covariate-adaptive randomization scheme is used,
where 
$$ \zeta_{jk}   =  \frac{(\pi_j \C_{kj} + \pi_k \C_{jk})^\top \Sigma^{-1}(\pi_j \C_{kj} + \pi_k \C_{jk}) }{\pi_j\pi_k (\pi_j+\pi_k)} +  \frac{(1-\pi_j-\pi_k)(\C_{jk} - \C_{kj})^\top \Sigma^{-1}(\C_{jk} - \C_{kj})  }{\pi_j+\pi_k},
$$
$\C_{jk} = {\rm Cov} \{ F_k(Y_j), W\}$, 
and $\C_{kj} = {\rm Cov} \{ F_j(Y_k), W\}$. \\ 
(iii) Assume (C) and $S_l$ is continuous, $l=j,k$. For censored outcome, 
$$ \sqrt{n} ( \hat \Delta_{jk}  - \Delta_{jk}  )
	\  \stackrel{d}{\to} \ N\left(0, \ \frac{{\rm Var}( {\mathscr D}_j )}{\pi_j} +   \frac{{\rm Var}( {\mathscr D}_k )}{\pi_k} - \eta_{jk} \right)  $$
 regardless of which covariate-adaptive randomization scheme is used,	where 
\begin{align*}
\eta_{jk} & = \frac{(\pi_j {\mathscr C}_{k} + \pi_k {\mathscr C}_{j})^\top \Sigma^{-1}(\pi_j {\mathscr C}_{k} + \pi_k {\mathscr C}_{j}) }{\pi_j\pi_k (\pi_j+\pi_k)}+  \frac{(1-\pi_j-\pi_k)({\mathscr C}_{j} - {\mathscr C}_{k})^\top \Sigma^{-1}({\mathscr C}_{j} - {\mathscr C}_{k})  }{(\pi_j+\pi_k)} ,\\
\mathscr D_l & =\int_0^\tau\frac{\int_t^\tau S_l(u)\,du}{P\{\min(Y_l,C_l)\ge t\mid A=l\}}
\,dM_l^0(t),
\end{align*}
${\mathscr C}_l = {\rm Cov} (W, {\mathscr D}_l)$, 
 % $M_l(t)   $ is defined in Theorem 1(ii), and $ M_l' (t) $ is equal to $M_l(t)$ with $N_l(t)$ replaced by $N_l'
 % (t) = I(A=l, \min(Y_l, C_l) \leq t)$,
{\color{black} $dM_l^0(t)$ is defined as $dM_l(t)$ in Theorem~1(ii), with $N_l(t)$ replaced by $I(Y_l \leq C_l,\, Y_l \leq t)$ and $Y_l(t)$ replaced by $I(Y_l \geq t,\, C_l \geq t)$.}
 %= I(U_j \leq t)- \int_0^t Y_j ( s) d  P\{ U_j \leq s \mid Y_j (s) = 1 \}$, 
\end{theorem}

Under simple randomization, $\tilde F_{j}^{-1}(p)$ and $ \tilde F_k^{-1}(p)$  are independent and $\sqrt{n} \{ \tilde F_{j}^{-1}(p)\!-\! \tilde F_k^{-1} (p)\!- \! F^{-1}_j(p)\!+ \!F_k^{-1}(p)\} \stackrel{d}{\to}  N\big(0, 
p(1-p)/ (\pi_j f_{jp}^2) + p(1-p)/ (\pi_k f_{kp}^2) \big) $, so the MELE $\hat F_{j}^{-1}(p)- \hat F_k^{-1} (p)$
has a guaranteed asymptotic efficiency gain.
Similarly, for the Wilcoxon two-sample statistic $\tilde\theta_{jk}$ defined at the end of Section~3.2,
$\sqrt{n} ( \tilde \theta_{jk} - \theta_{jk} )  \stackrel{d}{\to}  N\big(0, 
{\rm Var}\{F_k(Y_j)\}/ \pi_j + {\rm Var}\{ F_j(Y_k)\} /\pi_k \big) $ \citep{Jiang2010}, so the MELE $\hat\theta_{jk}$ also has a guaranteed asymptotic efficiency gain.
Finally, 
$ \sqrt{n} ( \tilde \Delta_{jk} - \Delta_{jk} )
\  \stackrel{d}{\to} \ N\big(0, {\rm Var}( {\mathscr D}_j )/\pi_j + {\rm Var}( {\mathscr D}_k )/\pi_k \big) $ \citep{claggett2022,sun2025}, so 
$\hat \Delta_{jk} $ likewise
has a guaranteed asymptotic efficiency gain. In each case, equality holds only in degenerate situations: for $J > 2$, when both arm-specific covariance vectors vanish (e.g., {$ \tilde\C_j=\tilde\C_k=0$} in the quantile treatment effect case); for $J = 2$, under the weaker condition that they cancel across arms (e.g., {$\pi_j \tilde\C_k+\pi_k\tilde\C_j=0$} in the quantile treatment effect case).

%{\color{blue} For $J>2$, equality holds if and only if the corresponding two covariance vectors are both zero (i.e. $\tilde \C_j=\tilde\C_k=0$ in the quantile treatment effect case). For $J=2$, equality holds if and only if  the corresponding two covariance vectors cancel between the two arms (i.e. $\pi_j\tilde \C_k+\pi_k\tilde\C_j=0$ in the quantile treatment effect case).}

%For the covariate adjusted  Wilcoxon two sample statistic $\hat\varpi_{jk}$ defined in the end of Section 3.1, we have the following corollary.  
 
% \begin{corollary}
% 	Assume (D) with  $Z$ included in $W$.  For the covariate adjusted Wilcoxon two sample statistic $\hat\varpi_{jk}$ defined in Section 3.1, 	 $$\sqrt{n} (\hat\varpi_{jk} - \varpi_{jk}) \ \stackrel{d}{\to} \ N\big(0, \, V_{jk} \big) $$ with 	
% \begin{align*}
% V_{jk}  & = \frac{ {\rm Var}\{F_k(Y_j) \}}{\pi_j} +  \frac{{\rm Var}\{F_j(Y_k) \}}{\pi_k} - \frac{(\pi_j c_{kj} + \pi_k c_{jk})^\top \Sigma^{-1}(\pi_j c_{kj} + \pi_k c_{jk}) }{\pi_j\pi_k (\pi_j+\pi_k)} \\
% & \quad - \frac{(1-\pi_j-\pi_k)(c_{jk} - c_{kj})^\top \Sigma^{-1}(c_{jk} - c_{kj})  }{(\pi_j+\pi_k)}, 
% \end{align*} 
% $~$  \\
% where $c_{jk} = {\rm Cov}\{ F_k (Y_j), W\}$ and $c_{kj} = {\rm Cov}\{ F_j (Y_k), W\}$. 
%\end{corollary}

\section{Variance Estimation and Confidence Intervals}

To assess the performance of an estimator of parameter $\theta$ or make asymptotic statistical inference about $\theta$, we need a consistent estimator of the asymptotic variance of estimator of $\theta$, regardless of which randomization scheme is used.
Once a consistent variance estimator is obtained, 
the Wald confidence interval with level $1- \alpha \in (0,1)$ 
%with two end points = the estimator of $\theta \pm \, z_{\alpha/2} \sqrt{\mbox{variance estimator}}$, where $z_a$ is the $1-a$ quantile of standard normal distribution, 
is asymptotically correct in the sense that  $\lim_{n\to \infty} P(\mbox{the interval covers $\theta$}) = 1-\alpha$. For the survival function $S_j(t)$, whose values lie in $(0,1)$, we use the log-transformed version of the Wald interval.

For non-censored outcome, 
a consistent  estimator of the asymptotic variance $V_{F_j}(y)$ of  $\hat F_j(y)$ in  Theorem 1(i) can be obtained by substituting $F_j(y)$ 
with  $\hat F_j(y)$, $\C_j(y)$ with  the sample covariance of $W_i$ and $I(Y_{iA_i} \leq y)$ over $i$ with  $A_i=j$, and 
 $ \Sigma$ with $\hat \Sigma =$ the sample covariance matrix of $W_i$, $i=1,...,n$,
regardless of which randomization scheme is used. 
%From the  MELE $\hat{\bar Y}_j$ of mean, a consistent  estimator of the asymptotic variance of $\hat{\bar Y}_j$ is 
%$\{ \hat V_j - (1-\pi_{j}) \hat C_j^\top \hat \Sigma^{-1} \hat C _j \}/ (\pi_j n )$,  regardless of which randomization scheme is used, where $ \hat C_j $ is the sample covariance of $X_i$ and $Y_{iA_i} $ over $i$ with  $A_i=j$
% = \sum_{i: A_i =j} \hat p_{ij} (\bX_i-  \bar X ) Y_{iA_i}  $$ and $\hat V_j$ is the sample variance of $Y_{iA_i}$ over $i$ with  $A_i=j$.        %Other consistent variance estimators can be similarly obtained. 
For the quantile estimator $\hat F_{j}^{-1}(p) $ or  $\hat F_{j}^{-1}(p)-\hat F_{k}^{-1}(p)$,  
a consistent variance estimator based on Theorems 1-3 involves consistently 
estimating the derivative $f_{jp}$ or derivatives $f_{jp}$ and $f_{kp}$, which is known to be difficult.
We run into the same issue for unadjusted quantile estimator and difference. 
We recommend using nonparametric bootstrap
variance estimation described in the Supplementary Material with some simulation
results. 
For the MELE $\hat\theta_{jk}$ considered in the end of Section 3.2, its consistent variance estimator can be obtained using Theorem 3(ii) with ${\rm Var} \{ F_j(Y_k)\}$ estimated by the sample variance based on $\hat F_j (Y_{iA_i})$, for $i$'s in the treatment group $k$, and {$\C_{kj}$ estimated by the sample covariance of $\hat F_j(Y_{iA_i}) $ and $W_i$, for $i$'s in the treatment group $k$.}

For censored outcome,  a consistent estimator of the asymptotic variance $V_{S_j}(t)$ in Theorem 1(ii) for $\hat S_j(t)$  is 
\begin{equation}
 {\hat V_{S_j}(t) = \frac{1}{n}\sum_{i=1}^n \frac{\hat S_j^2(t)}{\color{black}{\pi_j^2}} \bigg\{ \! \int_0^t \frac{d  \hat M_{ij}(s)}{\bar{Y}_j (s)} \bigg\}^2 - \frac{(1-\pi_j) \hat \Gamma_j(t)^\top \hat \Sigma^{-1} \hat\Gamma_j(t)}{\pi_j},}  \label{v1}
\end{equation}
where $ d \hat M_{ij}(t)  =  d N_{ij} (t) -  Y_{ij} (t) d \hat \Lambda_j(t)$,
$d \hat \Lambda_j(t)$ is given in \eqref{AKM}, $\bar Y_j (t) = n_j^{-1} \sum_{i: A_i =j} Y_{ij}(t)$, and
$ \hat\Gamma_j(t)  = \hat S_j(t)\sum_{i: A_i =j} (W_i-\bar W_j) \int_0^t  d\hat M_{ij}(s)/ \sum_{i': A_{i'} =j} Y_{i'j}(s). $
%$\hat M_j(t) = I(T_j \leq t, T_j \leq C_j)  + I(\min(T_j, C_j) \geq t)\log \hat S_j(t)$,
A consistent variance estimator for MELE $\hat \Delta_{jk}$ defined in the end of Section 3.3 and given in Theorem 3(iii)  is 
\[ 
	\hat {\mathscr V}_{jk}  = \frac{\hat {\mathscr V}_j}{\pi_j}  +\frac{\hat {\mathscr V}_k}{\pi_k} - \frac{(\pi_j \hat{\mathscr C}_{k} + \pi_k \hat{\mathscr C}_{j})^\top \hat\Sigma^{-1}(\pi_j \hat{\mathscr C}_{k} + \pi_k\hat{\mathscr C}_{j}) }{\pi_j\pi_k (\pi_j+\pi_k)} - \frac{(1-\pi_j-\pi_k)(\hat{\mathscr C}_{j} - \hat{\mathscr C}_{k})^\top \hat\Sigma^{-1}(\hat{\mathscr C}_{j} - \hat{\mathscr C}_{k})  }{(\pi_j+\pi_k)},  \]
where $\hat {\mathscr V}_j$ is the sample variance of 
{\color{black} 
\[
\hat{\mathscr D}_{ij} = \int_0^\tau \frac{\int_t^\tau \hat S_j(u)du}{ \bar Y_{j}(t)} d \hat M_{ij}^0(t),  \text{ for } A_i=j,
\]
 $d\hat M_{ij}^0(t)$ is defined as $d\hat M_{ij}(t)$, with $N_{ij}(t)$ replaced by $I(Y_{ij} \leq C_{ij},\, Y_{ij} \leq t)$ and $Y_{ij}(t)$ replaced by $I(Y_{ij} \geq t,\, C_{ij} \geq t)$,} 
and
$\hat{\mathscr C}_j$ is the sample covariance of $W_i$ and 
$\hat {\mathscr D}_{ij}$, $i=1,...,n$, $A_i=j$. 

%For the unadjusted $\tilde \U_j - \tilde \U_k$ based on Kaplan-Meier, the variance estimator is $\frac{\hat {\mathscr V}_j}{\pi_j}  +\frac{\hat {\mathscr V}_k}{\pi_k} $. 

All the results given here are valid 
regardless of which randomization scheme is used.
%The performance of  variance estimators is examined in the simulation  (Section 6).

\section{Simulation}

 We focus on right-censored time-to-event data and  
 consider a simulation to evaluate and compare the finite sample performances of our proposed MELE  of $S_j(t)$ given in \eqref{AKM}, the AGEE  in \cite{zhang2015robust} with the same covariate adjustment, and the Kaplan-Meier estimator in \eqref{KM} without covariate adjustment. 
Our simulation setting is the same  as  scenario 2 in \cite{zhang2015robust}, where 
$J=2$ with  $\pi_1=\pi_2 = 1/2$, sample size $n=200$, a single baseline covariate $X \sim N(0,1)$,  the  
life time outcome $Y_j $ conditioned on $X$ is distributed as $\log N\big( X + 0.3 (j-1), 1\big)$, and the censoring time $C_j \sim $ uniform(0, 7.5)$+ 0.7 (j-1)$, independent of $X$ and $Y_j$, $j=1,2$.
For assigning treatments, 
we employ simple randomization and covariate-adaptive randomization, the stratified permuted block of size 4, 
where the strata are formed by $Z = X$ discretized into four categories with equal probabilities based on the quartiles. 
%In case II, $Z$ has two components, $B$ and $W$ discretized into four categories with equal probabilities based on the quartiles.
The adjusted covariate $W$ used in (\ref{like}) is {\color{black} $(X_c, X_c^2, X_c^3)^\top$ under simple randomization and $(Z, X_c, X_c^2, X_c^3)^\top$ under stratified permuted block randomization,} with $X_c$ being $X$ truncated at $-5$ and 5. 
%In case II, $X= (1, Z, W)^\top$ with $W$ truncated at $-5$ and 5. 

For variance estimation, 
we use $\hat V_{S_j}(t)$  given by (\ref{v1}) for the proposed MELE.
For the AGEE, we use the variance estimator in  \cite{zhang2015robust}. 
 For the Kaplan-Meier estimator, we used the first term in  $\hat V_{S_j}(t)$ in (\ref{v1}), which is a consistent estimator under simple randomization, but overestimates under covariate-adaptive randomization such as the stratified permuted block in the simulation. 

Under the same simulation setting, we also evaluate the finite sample performances of the MELE $\hat \Delta_{21}$ proposed in the end of Section 3.3 and the unadjusted $\tilde \Delta_{21}$ (Kaplan-Meier), for estimating the difference 
$\Delta_{21}$  as the treatment effect. The unadjusted $\tilde \Delta_{21}$ and its variance estimator are computed using the R package survRM2. 
Variance estimation for $\hat\Delta_{21}$ is done using the results given in Section 5.

Based on 2,000 simulation replications, 
Table 1 gives the  results for estimating  survival function $S_j(t)$ for several $t$'s and $j=1,2$, and $\Delta_{21}$ with $\tau =3$:
the average of bias (AB),  standard deviation (SD), average of estimated SD (SE), coverage probability (CP) of 95\% Wald confidence interval,  and relative efficiency (RE) defined as  (SD$^2$ of unadjusted)/SD$^2$.

The following is a summary of the simulation results in Table 1. 
\begin{enumerate}
	\item All estimators have negligible average bias, which supports their asymptotic validity. 
	\item
In terms of simulation SD, 	the covariate adjusted estimators are comparable and both are more efficient than the unadjusted estimator under simple randomization, with  RE up to 1.21 for estimating the survival function and 1.66 for estimating the restricted mean survival time difference.
Under stratified permuted block randomization, the SD of Kaplan-Meier estimator is comparable with those of MELE and AGEE, confirming  the discussion (point \#3 in Section~4) when $Z$ provides most information contained in $W$.
\item The variance estimators for the MELE proposed in Section 5 and by \cite{zhang2015robust} perform well in terms of SE and CP, demonstrating their validity and confirming that they can be applied universally across randomization schemes. 
\item For the Kaplan-Meier estimator, the standard variance estimator works only under simple randomization and can overestimate under covariate-adaptive randomization. This confirms that, without covariate adjustment in estimation, variance estimation and inference using the Kaplan-Meier method are not valid under covariate-adaptive randomization. 
\item The AGEE estimator of \cite{zhang2015robust} is  not always a monotone function.
In 2,000 simulation runs, 2.60\% under simple randomization and 2.75\% under stratified permuted block randomization resulted in non-monotone AGEE curves (Table 1). Our MELE is always monotone by construction.
\end{enumerate}

{In the Supplementary Material, we also report simulation results under scenario 3 of \cite{zhang2015robust} with a multivariate $X$, as well as results for non-censored outcomes under simple randomization that evaluate and compare the finite-sample performance of the distribution estimators \eqref{fhatu}-\eqref{fhata} (covariate-adjusted and unadjusted) and the corresponding quantile estimators.}

\section{Application}

We apply the proposed method to real data from the SURPASS-4 study, a randomized, open-label, parallel-group, phase 3 trial assessing efficacy and safety of tirzepatide versus insulin glargine in adults with type 2 diabetes and high cardiovascular 
risk inadequately controlled on oral glucose-lowering medications \citep{del2021tirzepatide}. 

In this example, we focus on a cardio-renal composite endpoint, defined as the time to the first occurrence of any of the following: myocardial infarction, stroke, hospitalization due to heart failure, coronary revascularization, $\geq$ 40\% decline from baseline in estimated Glomerular Filtration Rate (eGFR), onset of macroalbuminuria, end-stage kidney disease, and all-cause death. We restrict our analysis to patients with kidney impairment (i.e. baseline eGFR  $<60 \text{ mL/min/}1.73\text{m}^2$). Among the $330$ patients, $173$ were randomized to tirzepatide (considered as treatment) and 157 were randomized to insulin glargine (considered as control), based on {stratified permuted block randomization} stratified by region (inside or outside of USA), baseline hemoglobin A1c (HbA1c) ($\leq 8.5$ \% or $>8.5$\%), and baseline use of sodium-glucose cotransporter-2 inhibitors {\color{black} (SGLT2i)}.

In the treatment group, 21 of  173 patients experienced at least one event during follow-up, corresponding to a 12.1\% event rate; in the control group, 41 of 157  experienced at least one event during follow-up,  corresponding to 26.1\% event rate. A Cox proportional hazards model including only the treatment indicator yielded a hazard ratio  of 0.50 with 95\% Wald confidence interval $(0.30, \, 0.85)$, indicating a substantial reduction in risk associated with tirzepatide.

\begin{figure}[ht]
    \centering
    \includegraphics[scale=0.6]{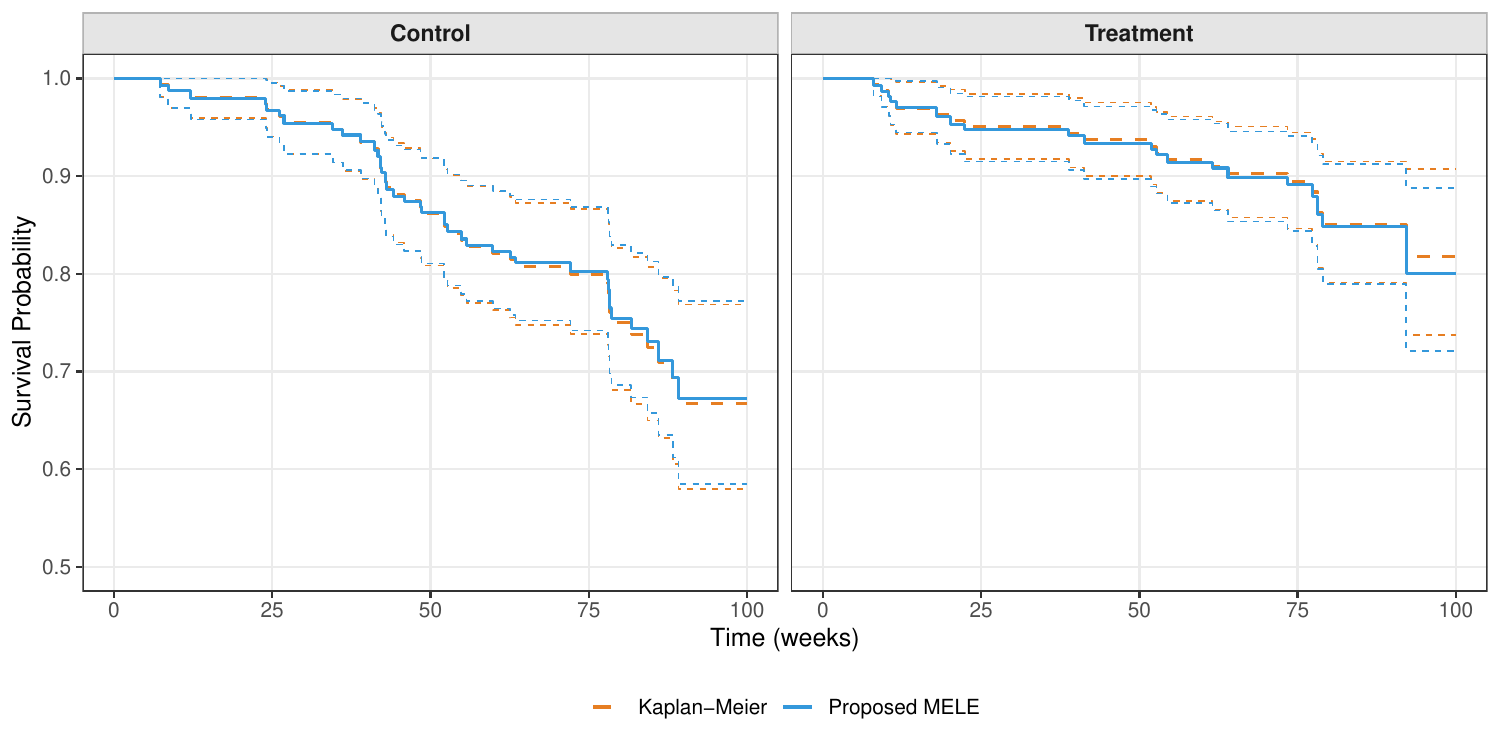}
    \caption{Estimated survival curves for the two groups using the Kaplan-Meier method and the proposed MELE, with the upper and lower curves representing the 95\% confidence bounds.}
    \label{fig:real data}
\end{figure}

The estimates of survival curves at selected weeks using the MELE 
(with baseline covariate adjustment for age, eGFR, HbA1c group, log-transformed urine albumin-creatinine ratio, region, {\color{black} and SGLT2i use})  and Kaplan-Meier are shown in {Figure \ref{fig:real data}}, together with 95\% {\color{black} log-transformed} confidence intervals.
{The values of estimates and standard errors  are given in Table S1 of the Supplementary Material. {\color{black} The Kaplan-Meier standard errors are computed using the variance estimator described in the simulation studies, which is valid under simple randomization but may be conservative under the stratified randomization used in this trial.} The reduction in SE$^2$ ranges from 0.8\% to 3.5\% for the control group and 0.7\% to 4.2\% for the treatment group, reflecting a consistent efficiency improvement from covariate adjustment.}

{We also estimate the restricted mean survival time (RMST) difference at $\tau=100$ weeks. The MELE yielded an estimated RMST difference of $5.35$ weeks (SE $=2.56$, 95\% CI: $0.32$ to $10.37$), compared to the Kaplan-Meier estimate of $5.84$ weeks (SE $= 2.64$, 95\% CI: $0.66$ to $11.01$). The SE$^2$ reduction for the RMST difference was $5.7\%$. Both methods yielded a statistically significant treatment benefit at the 5\% level, with the MELE providing a narrower confidence interval.}

%We consider real data in  the SURPASS-4  study of type 2 diabetes and increased cardiovascular risk   ({\red Lancet, 398(10313), 1811-1824}), a randomized, open-label, parallel-group, phase 3 trial. There two treatments, tirzepatide (considered as control) and insulin glargine  (considered as treatment) and  $T_j$ is a cardio-renal composite endpoint (myocardial infarction, stroke, heart failure, coronary revascularization, 40\% {\red eGFR} decline, onset of macroalbuminuria, end-stage kidney disease, and all-cause death). We focused on a subgroup with baseline eGFR $< 60$ with 173 and 157 patients under the treatment control groups, respectively.  The baseline covariates are age, eGFR, log-{\red UACR, HbA1c} group ($\leq 8.5\%$ or $>8.5\%$), and region (in or outside of USA). 

\begin{table}[hb]
\small
	\centering
	\caption{Simulation results based on 2,000 replications with $n=200$: AB = average bias, SD = standard deviation, SE = average of estimated SD, CP = coverage probability of 95\% Wald confidence interval, and RE = (SD$^2$ of Kaplan-Meier estimator) / SD$^2$} 
	\begin{tabular}{cccccccccc}
		\hline
		&  \multicolumn{3}{c}{Estimand} && \multicolumn{5}{c}{Performance of estimation} \\
		Randomization & $j$ & $t$ & $S_j(t)$ & \multicolumn{1}{c}{Estimator} & AB & SD & SE & CP & RE \\ \hline
		Simple & 1 & 1 & 0.500 & Kaplan-Meier & 0.000 & 0.052 & 0.052 & 0.949 &  \\
		 &  &  &  & Proposed MELE & 0.001 & 0.048 & 0.047 & 0.940 & 1.17 \\
		 &  &  &  & Zhang's AGEE & 0.001 & 0.047 & 0.047 & 0.944 & 1.19 \\
		 &  & 2 & 0.312 & Kaplan-Meier & 0.000 & 0.051 & 0.050 & 0.945 &  \\
		 &  &  &  & Proposed MELE & 0.000 & 0.047 & 0.045 & 0.942 & 1.15 \\
		 &  &  &  & Zhang's AGEE & 0.002 & 0.047 & 0.046 & 0.944 & 1.16 \\
		 &  & 3 & 0.219 & Kaplan-Meier & 0.000 & 0.047 & 0.047 & 0.952 &  \\
		 &  &  &  & Proposed MELE & 0.000 & 0.045 & 0.042 & 0.938 & 1.08 \\
		 &  &  &  & Zhang's AGEE & 0.002 & 0.045 & 0.043 & 0.938 & 1.08 \\
		 & 2 & 1 & 0.584 & Kaplan-Meier & 0.000 & 0.051 & 0.049 & 0.940 &  \\
		 &  &  &  & Proposed MELE & -0.001 & 0.047 & 0.044 & 0.937 & 1.19 \\
		 &  &  &  & Zhang's AGEE & -0.001 & 0.047 & 0.044 & 0.942 & 1.21 \\
		 &  & 2 & 0.391 & Kaplan-Meier & 0.000 & 0.050 & 0.050 & 0.954 &  \\
		 &  &  &  & Proposed MELE & 0.000 & 0.046 & 0.045 & 0.952 & 1.20 \\
		 &  &  &  & Zhang's AGEE & 0.001 & 0.046 & 0.045 & 0.955 & 1.21 \\
		 &  & 3 & 0.286 & Kaplan-Meier & 0.000 & 0.049 & 0.049 & 0.951 &  \\
		 &  &  &  & Proposed MELE & -0.001 & 0.046 & 0.044 & 0.945 & 1.14 \\
		 &  &  &  & Zhang's AGEE & 0.001 & 0.046 & 0.044 & 0.945 & 1.14 \\
		\cline{5-10}
		&&&& \multicolumn{6}{c}{Zhang's AGEE has 2.60\% non-monotone estimated curves} \\ \cline{5-10}
		&\multicolumn{3}{c}{$\Delta_{21} \ (\tau = 3)$} & Kaplan-Meier & 0.000 & 0.158 & 0.161 & 0.955 &  \\
		&\multicolumn{3}{c}{$0.216$} & Proposed MELE & 0.000 & 0.123 & 0.121 & 0.952 & 1.66 \\
		\hline
		Stratified & 1 & 1 & 0.500 & Kaplan-Meier & 0.000 & 0.048 & 0.052 & 0.962 &  \\
		permuted &  &  &  & Proposed MELE & 0.000 & 0.048 & 0.046 & 0.938 &  \\
		block &  &  &  & Zhang's AGEE & 0.000 & 0.048 & 0.047 & 0.940 &  \\
		 &  & 2 & 0.312 & Kaplan-Meier & 0.000 & 0.046 & 0.050 & 0.966 &  \\
		 &  &  &  & Proposed MELE & -0.001 & 0.047 & 0.045 & 0.946 &  \\
		 &  &  &  & Zhang's AGEE & 0.001 & 0.046 & 0.045 & 0.947 &  \\
		 &  & 3 & 0.219 & Kaplan-Meier & 0.000 & 0.045 & 0.047 & 0.958 &  \\
		 &  &  &  & Proposed MELE & -0.001 & 0.045 & 0.042 & 0.931 &  \\
		 &  &  &  & Zhang's AGEE & 0.001 & 0.045 & 0.042 & 0.932 &  \\
		 & 2 & 1 & 0.584 & Kaplan-Meier & -0.001 & 0.046 & 0.049 & 0.974 &  \\
		 &  &  &  & Proposed MELE & 0.000 & 0.045 & 0.044 & 0.946 &  \\
		 &  &  &  & Zhang's AGEE & -0.001 & 0.045 & 0.044 & 0.946 &  \\
		 &  & 2 & 0.391 & Kaplan-Meier & -0.001 & 0.046 & 0.050 & 0.970 &  \\
		 &  &  &  & Proposed MELE & -0.002 & 0.046 & 0.045 & 0.950 &  \\
		 &  &  &  & Zhang's AGEE & -0.001 & 0.045 & 0.045 & 0.954 &  \\
		 &  & 3 & 0.286 & Kaplan-Meier & -0.001 & 0.046 & 0.048 & 0.966 &  \\
		 &  &  &  & Proposed MELE & -0.001 & 0.046 & 0.043 & 0.939 &  \\
		 &  &  &  & Zhang's AGEE & 0.000 & 0.046 & 0.044 & 0.936 &  \\
		\cline{5-10}
		&&&& \multicolumn{6}{c}{Zhang's AGEE has 2.75\% non-monotone estimated curves} \\ \cline{5-10}
		&\multicolumn{3}{c}{$\Delta_{21} \ (\tau = 3)$} & Kaplan-Meier & -0.003 & 0.125 & 0.161 & 0.986 &  \\
		&\multicolumn{3}{c}{$0.216$} & Proposed MELE & -0.003 & 0.123 & 0.120 & 0.941 &  \\

		\hline
	\end{tabular}
\end{table}

\clearpage
\appendix
\setcounter{section}{0}
\renewcommand{\thesection}{S\arabic{section}}
\renewcommand{\theequation}{S\arabic{equation}}
\renewcommand{\thetable}{S\arabic{table}}
\renewcommand{\thefigure}{S\arabic{figure}}
\renewcommand{\thetheorem}{S\arabic{theorem}}
\renewcommand{\thelemma}{S\arabic{lemma}}

\begin{center}
{\Large\bf Supplementary Material}\\[6pt]
{\large Shape-Preserving Covariate Adjustment via Empirical Likelihood in Randomized Experiments}
\end{center}

\section{The bootstrap variance estimator for MELE of quantiles based on non-censored outcomes} 

% We consider the variance estimator 
% by bootstrapping outcomes and covariates within treatment group $j$.
{We use the ordinary nonparametric bootstrap to estimate the variance of the proposed MELE quantile estimators.
Specifically, draw
\(\{(A_i^*,Y_i^*,W_i^*):i=1,\ldots,n\}\) as a simple random sample with replacement from \(\{(A_i,Y_i,W_i):i=1,\ldots,n\}\), where $Y_i=Y_{iA_i}$. Let $n_j^*=\sum_{i=1}^n I(A_i^*=j)$. Then we recompute \(\muxh^*,\;(\hat\lambda_1^*,\ldots,\hat\lambda_J^*),\;\hat p_{ij}^*\) from the same equations applied to the bootstrap data, and hence $\hat F_j^*$:
% Specifically, let $\{(Y_{ij}^*, W_{i}^*), \, \mbox{all $i$ with $ A_i=j$}\}$ be a simple random sample with replacement  from $\{(Y_{ij},W_{i}), \,\mbox{all $i$ with $ A_i=j$}\}$, the data under $j$th treatment group, 
% and let 
 
$$	\hat F_j ^*(y) =
\sum_{i: A_i^*=j}  \hat p_{ij}^*  I(Y_{i}^* \leq y), \qquad - \infty < y < \infty, \quad j=1,...,J, $$
where   
$$
\hat p_{ij}^* = \frac{I(A_i^*=j)}{n_j^* + \hat \lambda_j^{*\top} (W_i^*  -  \muxh^* )}, \quad 	\sum_{j=1}^{J} \hat\lambda_j^* = 0 , \quad  \sum_{i=1}^n  \frac{ I(A_i^*=j)(W_i^* - \muxh^*  )}{n_j^*+ \hat\lambda_j^{*\top} (W_i^* - \muxh^*    )} = 0,
\quad j=1,...,J  
$$
%The  $p$-quantile of $	\hat F_j ^*$ is $\hat F_{j}^{*-1}(p)= \inf \{ y\! :\, \hat F_j^*(y) \geq p\}$.
The bootstrap variance estimator for $\hat F_{j}^{-1}(p)$ is 
${\rm Var}^* \{ \hat F_{j}^{*-1}(p)\}$ and for $\hat F_{j}^{-1}(p) - \hat F_{k}^{-1}(p)$ 
is ${\rm Var}^* \{ \hat F_{j}^{*-1}(p)-\hat F_{k}^{*-1}(p)\}$, where ${\rm Var}^* $ is the variance with respect to  bootstrap sampling conditional on { \(\{(A_i,Y_{iA_i},W_i):i=1,\ldots,n\}.\)}
% $\{(Y_{ij},W_{i}), \, \mbox{all $i$ with $ A_i=j$}\}$.
In applications, we use the bootstrap Monte Carlo, i.e., compute $\hat \theta^*$'s based on  independent
$B$ bootstrap datasets,  where $\hat \theta^* = \hat F_{j}^{*-1}(p)$ or $ \hat F_{j}^{*-1}(p)- \hat F_{k}^{*-1}(p)$, and
then approximate  ${\rm Var}^* (\hat \theta^*)$ by the sample variance of $\hat \theta^*$'s. 

The asymptotic distribution of the proposed MELE is invariant to the randomization scheme under condition (D). Therefore the same bootstrap variance estimator is valid under simple randomization and covariate-adaptive randomization satisfying (D). The bootstrap variance estimator for the augmented estimator based on \(\hat F_l^{\rm A}\) can be obtained similarly by recomputing the augmentation coefficient in each bootstrap sample. For the unadjusted quantile estimators, the ordinary bootstrap works under simple randomization, while under covariate-adaptive randomization their limiting variance may depend on the randomization scheme.}

%\begin{lemma}
%	The $\hat \lambda_j$ given in (4) satisfies 
%	\[ \frac{\hat{\lambda}_j}{n} = \Sigma^{-1} \bar u_j +  \frac{o_p(1)}{\sqrt{n}},	\quad	\qquad \bar u_j = \frac{1}{n}\sum_{i=1}^{n} \left\{I(A_i=j) - \pi_j\right\} (X_{i} - \mux ). 	\]
%\end{lemma}

\section{Proof of Lemma \ref{lemma1}}
%From () in Section 3, 
%\begin{equation}
%\hat p_{ij} = \frac{I(A_i=j)}{n_j +\hat \lambda_j^\top ( X_i -  \muxh )} ,  \quad 	\sum_{i=1}^{n}\frac{I(A_i=j) ( X_i -  \muxh )}{n_j + \hat{\lambda}_j^\top ( X_i -  \muxh )} =0, \quad  j=1,...,J,  \quad i=1,...,n.  \label{e1}
%%\end{equation}
 
For two sequences $\{\xi_n\}$ and $\{a_n\}$, write $\xi_n=O_p(a_n)$ if $\xi_n/a_n$ is bounded in probability and $\xi_n=o_p(a_n)$ if $\xi_n/a_n=o_p(1)$.

We first prove several useful results under Condition (D). Since $Z$ has finite support and $n_{zj}-\pi_jn_z=O_p(n_z^{1/2})$, $n_j/n=\pi_j+O_p(n^{-1/2}).$
Let $m_z=E(W-\mux\mid Z=z)$. Then  
\[
\sum_{i:A_i=j}(W_i-\mux)=\sum_z n_{zj}m_z+\sum_z\sum_{i:Z_i=z,A_i=j}\{W_i-\mux-m_z\}.
\]  
The first term is $O_p(n^{1/2})$ by Condition (D) and $E m_Z=0$, while the second is $O_p(n^{1/2})$ by Chebyshev's inequality. Thus, with $a_n=\max_j\|\bar W_j-\mux\|$ and $M_n=\max_i\|W_i-\mux\|$, we have $a_n=O_p(n^{-1/2})$. Similarly, 
\[
    \frac{1}{n_j}\sum_{i:A_i=j}(W_i-\mux)(W_i-\mux)^\top
    = \Sigma+o_p(1) .
\]
Also, $M_n=o_p(n^{1/2})$ by lemma 11.2 in \cite{owen2001}. Hence $a_nM_n=o_p(1)$.

We next prove the preliminary rates $\max_j\|\hat\lambda_j\|/n=O_p(n^{-1/2})$ and $\muxh-\mux=O_p(n^{-1/2})$. First consider an empirical likelihood problem where the $\mux$ in \eqref{like} is the true mean, and let $\alpha_j$ be its multiplier satisfying 
\[
0=\frac1{n_j}\sum_{i:A_i=j}\frac{W_i-\mux}{1+\alpha_j^\top(W_i-\mux)}.
\]
Algebra gives
\[
\bar W_j-\mux=\frac{1}{n_j}\sum_{i:A_i=j}\frac{\alpha_j^\top (W_i-\mux)}{1+\alpha_j^\top (W_i-\mux)}(W_i-\mux). 
\]
Let $u_j=\alpha_j/\|\alpha_j\|$ if $\alpha_j\neq 0.$ Multiplying on the both sides by $u_j$ yields 
\[
u_j^\top(\bar W_j-\mux)=\|\alpha_j\|\frac1{n_j}\sum_{i:A_i=j}\frac{\{u_j^\top(W_i-\mux)\}^2}{1+\|\alpha_j\|u_j^\top(W_i-\mux)}. 
\]
Note that the denominators are positive and are bounded as $1+\|\alpha_j\|u_j^\top(W_i-\mux)\leq 1+\|\alpha_j\|\|W_i-\mux\|\leq 1+\|\alpha_j\|M_n$. Therefore, 
\[
\|\bar W_j-\mux\|\geq u_j^\top (\bar W_j-\mux)\geq \frac{\|\alpha_j\|}{1+\|\alpha_j\|M_n}u_j^\top\left\{\frac{1}{n_j}\sum_{i:A_i=j}(W_i-\mux)(W_i-\mux)^\top\right\}u_j. 
\]
Since the sample second-moment matrix has eigenvalues bounded away from zero with probability tending to one, there exists a constant $c>0$ such that $u_j^\top \Sigma u_j\geq c^{-1}$ uniformly over $j$. Thus $\|\alpha_j\|\leq c\,a_n(1+\|\alpha_j\|M_n)$. Because $a_nM_n=o_p(1)$, it follows that $\max_j\|\alpha_j\|=O_p(a_n)=O_p(n^{-1/2})$.

Let $t_{ij}=\alpha_j^\top (W_i-\mux)$. Then multiplying on the both sides of the score equation by $\alpha_j$ gives $\sum_{i:A_i=j}t_{ij}/(1+t_{ij})=0$, which implies 
$$\sum_{i:A_i=j}\log(1+t_{ij})=\sum_{i:A_i=j}\{\log(1+t_{ij})-\frac{t_{ij}}{1+t_{ij}}\}. $$
Define $h(t)=\log(1+t)-t/(1+t)$. Some calculations show $h(t)\geq 0$ for $t>-1$. Using Taylor expansion around $0$ gives $h(t)=t^2/2+O(t^3)$, which means there exists a constant $C>0$ such that $h(t)\leq Ct^2$ when $t$ is in a neighborhood of zero. Since
$\max_{i,j}|t_{ij}|\le M_n\max_j\|\alpha_j\|=o_p(1)$, 
\begin{align*}
0&\leq\sum_{i:A_i=j}\log(1+t_{ij})\\
&\leq C\,\sum_{i:A_i=j} t_{ij}^2\\
&= Cn_j\alpha_j^\top\{\frac{1}{n_j}\sum_{i:A_i=j}(W_i-\mux)(W_i-\mux)^\top\}\alpha_j\\
&\leq C\, n \|\alpha_j\|^2\,O_p(1)\\
&\leq C\, c^2\,n\, \, a_n^2\\
&=C_0na_n^2,
\end{align*}

\noindent where $C_0:=Cc^2.$

Now return to the fitted target $\muxh$. Since $\muxh$ maximizes the profile empirical likelihood, the fitted total loss is no larger than the fixed target loss. Therefore, with $s_{ij}=\hat\lambda_j^\top(W_i-\muxh)/n_j$, for every arm $j$, 
\[
0\leq \sum_{i:A_i=j}\left\{\log(1+s_{ij})-\frac{s_{ij}}{1+s_{ij}}\right\}\leq C_0na_n^2. 
\]
Using $\log(1+s)-s/(1+s)\geq s^2/(8+8|s|^2)$, we obtain 
\[
\sum_{i:A_i=j}\frac{s_{ij}^2}{(1+|s_{ij}|)^2}\leq 8C_0na_n^2. 
\]
Because $\muxh=\sum_{i:A_i=j}\hat p_{ij}W_i$ and $\sum_{i:A_i=j}\hat p_{ij}=1$, $\hat p_{ij}\ge0$, we have $\|\muxh-\mux\|\leq \max_{i:A_i=j}\|W_i-\mux\|\leq M_n$. And hence $\|W_i-\muxh\|\leq \|W_i-\mux\|+\|\mux-\muxh\|\leq 2M_n$. So $\|s_{ij}\|\leq 2M_n\|\hat\lambda_j\|/n_j$. Together, we have 
\[
\frac{\sum_{i:A_i=j}s_{ij}^2}{(1+2M_n\|\hat\lambda_j\|n_j^{-1})^2}\leq 8C_0na_n^2. 
\]
Noticing that 
\[
\sum_{i:A_i=j}s_{ij}^2=n_j\frac{\hat\lambda_j^\top}{n_j}\left\{\frac{1}{n_j}\sum_{i:A_i=j}(W_i-\muxh)(W_i-\muxh)^\top\right\}\frac{\hat\lambda_j}{n_j} 
\]
and following the derivations above, we obtain that there exists a constant $c_1^{-1}$ such that
\(
\sum_{i:A_i=j}s_{ij}^2\geq c_1\|\hat\lambda_j\|^2/{n_j}\geq c_1\|\hat\lambda_j\|^2/n.
\)
Therefore, by taking the square root and absorbing all constants into $C$, 
\[
\frac{\|\hat\lambda_j\|}{n}\leq Ca_n(1+2M_n\frac{\|\hat\lambda_j\|}{n}),
\]
which is equivalent to 
\[
\frac{\|\hat\lambda_j\|}{n}\leq \frac{Ca_n}{1-2Ca_nM_n}. 
\]
Using $a_nM_n=o_p(1)$, we have $\|\hat\lambda_j\|/n_j\leq Ca_n$ with probability approaching 1. Recalling that $a_n=O_p(n^{-1/2})$ completes the proof of $\hat\lambda_j/n = O_p(n^{-1/2})$. It remains to show $\muxh-\mux=O_p(n^{-1/2})$. The fitted score equation gives the exact 
\[
\bar W_j-\muxh=(\frac{1}{n_j}\sum_{i:A_i=j}\frac{(W_i-\muxh)(W_i-\muxh)^\top}{1+\frac{\hat\lambda^\top_j}{n_j}(W_i-\muxh)})\frac{\hat\lambda_j}{n_j}. 
\]
Some calculations yield $\|\bar W_j-\muxh\|\leq 2M_n$ and there exists a constant $C_1>0$ such that 
\[ 
\|\bar W_j-\muxh\|\leq C_1(\frac{\|\hat\lambda_j\|}{n}+\frac{\|\hat\lambda_j\|}{n}M_n\|\bar W_j-\muxh\|). 
\]
Because $M_n\max_j\|\hat\lambda_j\|/n_j=o_p(1)$, we have $\|\bar W_j-\muxh\|\leq C\|\hat\lambda_j\|/n_j=O_p(n^{-1/2})$ with probability approaching 1. Together with $\|\bar W_j-\mux\|=O_p(n^{-1/2})$, this yields $\|\muxh-\mux\|=O_p(n^{-1/2}).$

We now complete the proof. With probability approaching one, $\max_{i,j}|\hat\lambda_j^\top(W_i-\muxh)|\leq n_j/2$ because $\|\hat\lambda_j\|=O_p(n^{1/2})$, $M_n=o_p(n^{1/2})$, and $\|\muxh-\mux\|=O_p(n^{-1/2})$. On this event, using 
\[
\frac{1}{n_j+x}=\frac{1}{n_j}-\frac{x}{n_j^2}+\frac{x^2}{n_j^2(n_j+x)} 
\]
with $x=\hat\lambda^\top_j(W_i-\muxh)$ in the $j$-th score equation gives 
\begin{equation}
    0=\frac{1}{n_j}\sum_{i:A_i=j} ( W_i -  \muxh ) - 
\frac{1}{n_j} \sum_{i:A_i=j} ( W_i -  \muxh )( W_i -  \muxh )^\top \, \frac{\hat{\lambda}_j}{n_j}+R_{nj},\tag{S1}
\label{S1} 
\end{equation}
where the remainder is 
\[
R_{nj} = \frac{1}{n_j^2}\sum_{i:A_i=j}\frac{[\hat\lambda_j^\top(W_i-\muxh)]^2(W_i-\muxh)}{n_j+\hat\lambda^\top_j(W_i-\muxh)}. 
\]
With probability approaching 1, as \(n_j+\hat\lambda_j^\top(W_i-\muxh)\ge n_j/2\), we bound the remainder as 
\[
\|R_{nj}\|\leq \frac{2}{n_j^3}\sum_{i:A_i=j}[\hat\lambda_j^\top(W_i-\muxh)]^2\|W_i-\muxh\|\leq\frac{2\|\hat\lambda_j\|^2}{n_j^3}\sum_{i:A_i=j}\|W_i-\muxh\|^3. 
\]
Given $E\|W\|^3<\infty$, it yields $\|R_{nj}\|\leq O_p(n^{-1})$. Using $\muxh-\mux=O_p(n^{-1/2})$ and the covariance convergence gives 
\[
0=\frac{1}{n}\sum_{i:A_i=j}(W_i-\muxh)-\Sigma\frac{\hat\lambda_j}{n}+o_p(n^{-1/2}). 
\]
The first term is 
$$n^{-1}\sum_{i:A_i=j}(W_i-\muxh)=\bar U_j+\pi_j(\bar W-\mux)-n_j(\muxh-\mux)/n=\bar U_j+\pi_j(\bar W-\mux)-\pi_j(\muxh-\mux)+o_p(n^{-1/2}). $$ 
Therefore for each $j$, 
\[
0=\bar U_j+\pi_j(\bar W-\mux)-\pi_j(\muxh-\mux)-\Sigma\frac{\hat\lambda_j}{n}+o_p(n^{-1/2}) 
\]
Summing over $j$ gives $\muxh=\bar W+o_p(n^{-1/2})$. Substituting this back into the above equation yields \(n^{-1}\hat\lambda_j=\Sigma^{-1}\bar U_j+o_p(n^{-1/2})\).

% By \cite{qinlawless1994}, $\hat \lambda_j  /n=  o_p(1)$, $j=1,...,J$.
% Note that 
% \[
% \frac{\partial}{\partial \lambda_j} \left\{ \frac{I(A_i = j)( W_i -  \muxh )}{n_j + \lambda_j^\top( W_i -  \muxh )}\right\} \bigg|_{\lambda_j = 0} = - \, \frac{ I(A_i = j) ( W_i -  \muxh ) ( W_i -  \muxh )^\top}{n_j^2} , \quad j=1,...,J.
% \]
% By \eqref{phat}, 
% and the algebraic identity $(n_j+x)^{-1}=n_j^{-1}-n_{j}^{-2}x+n_j^{-2}(n_j+x)^{-1}x^2$ with $x = \hat\lambda_j^\top(W_i-\muxh)$, $j=1,...,J$, we have
% \[
% \frac{1}{n_j}\sum_{i=1}^{n} I(A_i=j) ( W_i -  \muxh ) - 
% \frac{1}{n_j^2} \sum_{i=1}^{n} I(A_i=j) ( W_i -  \muxh )( W_i -  \muxh )^\top \, \hat{\lambda}_j+R_n^{(j)}, 
% \]
% where the remainder is
% \[
% R_n^{(j)} = \frac{1}{n_j^2}\sum_{i=1}^nI(A_i=j)\hat p_{ij}[\hat\lambda_j^\top(W_i-\muxh)]^2(W_i-\muxh).
% \]
% This together with   the law of large numbers imply
% \begin{equation}
% \frac{1}{n}\sum_{i=1}^{n} I(A_i=j)W_{i} - \pi_j \muxh -
% \Sigma \, \frac{\hat{\lambda}_j}{n} +\frac{o_p(1)}{\sqrt{n}}=0, \quad j=1,...,J. \tag{S1}
% \label{e3}
% \end{equation}
% Summing over $j$ in the left side of  (\ref{e3}) and using the facts that $\sum_{j=1}^J \pi_j =1$,  $\sum_{j=1}^J \hat \lambda_j = 0$, and $\sum_{j=1}^J I(A_i=j) =1$, we obtain that $
% \muxh   = \bar W + n^{-1/2}o_p(1) $, the first result. 

% The second result follows from (\ref{e3}) with $\muxh$ substituted by $
% \bar W + n^{-1/2}o_p(1) $ and the result
% \vspace{2mm} $n^{-1} \sum_{i=1}^n \{ I(A_i= j) - \pi_j\}(\bar W-\ \mux) = n^{-1/2}o_p(1) $.
%\pagebreak 

\section{Proof of Theorem 1(i)}
By (2), Taylor's expansion, the law of large numbers, and Lemma 1, 
\begin{align*}
\sum_{i: A_i =j} I(Y_{ij}\leq y) \bigg( \hat p_{ij} - \frac{1}{n_j} \bigg) & = - \sum_{i: A_i =j} I(Y_{iA_i}\leq y) \frac{(W_i - \mux)^\top \, \hat\lambda_j }{n_j^2} + \frac{o_p(1)}{\sqrt{n_j}} \\
& = - \, \frac{{\mathbb C}_j(y)^\top \Sigma^{-1} \bar u_j }{\pi_j}+ \frac{o_p(1)}{\sqrt{n}},
\end{align*}

\noindent 
where $\bar u_j = n^{-1} \sum_{i=1}^n \{ I(A_i= j) - \pi_j\}(W_i-\ \mux) $. Hence,  
\begin{align*}
\hat F_j (y) - F_j(y) & =   \tilde F_j(y) - F_j(y) + \sum_{i: A_i =j} I(Y_{iA_i}\leq y) \bigg( \hat p_{ij} - \frac{1}{n_j} \bigg)  \\
& =  \frac{1}{n} \sum_{i=1}^n  \frac{I(A_i=j)  \{ I(Y_{ij}\leq y) - F_j(y)\}}{\pi_j}  - \frac{{\mathbb C} _j(y)^\top  \Sigma^{-1} \bar u_j }{\pi_j} +\frac{o_p(1)}{\sqrt{n}} \\
& = \frac{1}{n} \sum_{i=1}^n \phi_{ij}(y) + \frac{o_p(1)}{\sqrt{n}}. 
\end{align*}

\noindent
This proves the first result in Theorem 1(i). 
Let $\C_j$ and $\V_j$ be covariance and variance conditioned on $n_j$. Then  

\begin{align*}
&\  \V_j \big\{ \hat F_j(y) - F_j(y)      - \pi_j^{-1}{\mathbb C}_j(y)^\top \Sigma^{-1}  \bar u_j  \big\}\\
= & \ \V_j \{  \hat F_j(y)  \}  -2  \, \C_j \big\{ \pi_j^{-1}{\mathbb C}_j(y)^\top\Sigma^{-1}  \bar u_j   , \hat F_j(y)\big\}   
+ \V_j\big\{ \pi_j^{-1}\C_j(y)^\top\Sigma^{-1}  \bar u_j  \big\}  \\
= & \  \frac{ F_j(y) \{1-F_j(y) \}}{n_j} - 2 \pi_j^{-1}\, \C_j(y)^\top \Sigma^{-1} \, \C_j \{  \bar \bu_j, \hat F_j(y)  \}  +  \pi_j^{-2} \C_j(y)^\top \Sigma^{-1} \V_j (\bar u_j )  \Sigma^{-1} \C_j(y)\\
= & \  \frac{ F_j(y) \{1-F_j(y) \}}{\pi_j n} -\frac{2(1-\pi_j)}{\pi_j n} \, \C_j(y)^\top\Sigma^{-1} \C_j (y) + \frac{1-\pi_j}{\pi_jn} \C_j(y)^\top  \Sigma^{-1} \C_j(y) +\frac{o_p(1)}{n}\\
= & \  \frac{ F_j(y) \{1-F_j(y) \}}{\pi_j n} - \frac{1-\pi_j}{\pi_jn } \, \C_j(y)^\top\Sigma^{-1} \C_j (y)+\frac{o_p(1)}{n} \\
= & \ \frac{V_{F_j}(y)}{n} + \frac{o_p(1)}{n}.
\end{align*}

\noindent
Hence, the convergence in distribution result in Theorem 1(i) 
 follows from   the central limit theorem  
\citep{ye2021better} applied to the sample mean $n^{-1} \sum_{i=1}^n \phi_{ij}$,  under covariate-adaptive randomization satisfying (D).

\section{Proof of Theorem 1(ii)} 
By (2), Taylor's expansion for $\hat\lambda_j/n$ at 0, Lemma 1, the law of large numbers, for
$s \leq \tau_j$,  
\begin{equation}
\begin{array}{l}
{\displaystyle\sum_{i=1}^n \hat p_{ij}  Y_{ij}(s) = \frac{1}{\pi_jn} \sum_{i=1}^n Y_{ij}(s)+o_p(1) = \frac{E\{ Y_{j}(s)\}}{\pi_j} +o_p(1)} , \vspace{2mm}\\
{\displaystyle \sum_{i=1}^n \hat{p}_{ij} dM_{ij}(s) 
= \frac{1}{n} \sum_{i=1}^n  \frac{\pi_j - ( W_i -  \muxh )^\top \hat\lambda_j  /n }{ \pi_j^2} dM_{ij}(s) + \frac{o_p(1)}{\sqrt{n}}. } \end{array} \tag{S2}
\label{e5}
\end{equation}
Under (C) and $t \leq \tau_j$,  $S_j(t)>0$ and
$E\{ Y_{j}(t)\}>0$. Hence, by the first result in (\ref{e5}), for $t \leq \tau$, with probability tending to 1 as $n \to \infty$, the estimator $\hat S_j(t)$ is well defined for $t \leq \tau$. 
From the argument in Chapter 3 in \cite{flemming1991} for simple randomization and in \cite{Wang:2021wg} for covariate-adaptive randomization,  $\widehat{S}_j(t)$ is consistent for $t \leq \tau$. 
Similar to the proof of Theorem 3.2.3 in \cite{flemming1991}, by the formulas for integration by parts and the differential of a reciprocal, we have
\begin{align*}
\widehat{S}_j(t)-S_j(t) & = -S_j(t) \int_0^t \frac{\widehat{S}_j(s-)}{S_j(s)} \left\{ \frac{\sum_{i=1}^n \hat p_{ij}  dN_{ij}(s)}{\sum_{i=1}^n \hat p_{ij}  Y_{ij}(s) } + d\log S_j(s) \right\}  \\
& = -\pi_jS_j(t)	  \int_0^t  \frac{1}{E\{Y_{j}(s)\}} \sum_{i=1}^n \hat p_{ij}  dM_{ij}(s) + \frac{o_p(1)}{\sqrt{n}}  \\
& = \frac{S_j(t)}{n} \left\{ -\sum_{i=1}^n \int_0^t \frac{dM_{ij}(s)}{E\{Y_{j}(s)\}}  +  \sum_{i=1}^n \int_0^t \frac{( W_i -  \muxh )^\top dM_{ij}(s)}{\pi_j E\{Y_{j}(s)\}} \frac{\hat\lambda_j}{n}  \right\} + \frac{o_p(1)}{\sqrt{n}} \\
%		The derivatives of the consistency for $\hat{S}_j(t)$ can be divided into two steps. 		First, show that for any fixed $u \in (0, \infty)$ such that as $n_j \to \infty$, $\sum_{i=1}^{n_j}{Y_{ij}}(u) \stackrel{P}{\to} \infty$, 		then $\sup_{0 \leqslant s \leqslant u} |\widehat{S}_j(s) - S_j(s)| \stackrel{P}{\to} 0$, as $n \to \infty$. 		Second, show that if $t \in (0, \infty]$ is such that for any $u < t$, $\sum_{i=1}^{n_j}{Y_{ij}}(u) \stackrel{P}{\to} \infty$, as $n \to \infty$. 		Then $\sup_{0 \leqslant s \leqslant t} |\widehat{S}_j(s) - S_j(s)| \stackrel{P}{\to} 0$, as $n \to \infty$.
&= -\frac{S_j(t)}{n} \sum_{i=1}^n \int_0^t \frac{dM_{ij}(s)}{ E\{ Y_{j}(s)\} } +  S_j(t)E \left\{\frac{W^\top}{\pi_j}\int_0^t \frac{ dM_{j}(s)}{E\{ Y_{j}(s)\}}  \right\} \frac{\hat{\lambda}_j}{n} + \frac{o_p(1)}{\sqrt{n}}\\
& = \frac{1}{n} \sum_{i=1}^n \left[ -S_j(t)\int_0^t \frac{dM_{ij}(s)}{ E\{ Y_{j}(s)\} } +    \frac{\left\{I(A_i=j) - \pi_j\right\} \Gamma_j(t)^\top  \Sigma^{-1}{(W_{i}-\mux })}{\pi_j}\right] + \frac{o_p(1)}{\sqrt{n}} \\
& = \frac{1}{n} \sum_{i=1}^n \varphi_{ij} (t) + \frac{o_p(1)}{\sqrt{n}},
\end{align*}

\noindent
where the second equality follows from the consistency of $\hat S_{j}(t)$, the first  result in (\ref{e5}), the definition of $M_{ij}(t)$, and the fact that the first term at the right side  multiplying by $\sqrt{n} $ is asymptotically normal, 
the third equality follows from the second result in (\ref{e5}), the fourth equality follows from the law of large numbers, and the fifth equality follows from Lemma 1. 
This proves the first result in Theorem 1(ii).

By the central limit theory for average of random variables under covariate-adaptive randomization \citep{Ye:2020ab}, we can obtain that $\sqrt{n}  \{ \hat S_j(t) - {S}_j(t) \} \stackrel{d}{\to} N(0, V_{S_j}(t))$, $j=1,...,J$,
where 
\begin{align*}
V_{S_j}(t) & = {\rm Var} \left[ S_j(t) \int_0^t\frac{dM_{j}(s)}{E\{ Y_{j}(s)\}} -    \frac{\left\{I(A=j) - \pi_j\right\} \Gamma_j(t)^\top  \Sigma^{-1}(W-\mux) }{\pi_j} \right] \\
& = {\rm Var} \left\{  S_j(t)\int_0^t\frac{dM_{j}(s)}{E\{ Y_{j}(s)\}} \right\}+ {\rm Var} 
\bigg[ \frac{\left\{I(A=j) - \pi_j\right\} \Gamma_j(t)^\top  \Sigma^{-1}(W-\mux)}{\pi_j} \bigg] \\
& \quad - 2  {\rm Cov} \bigg[ S_j(t)\int_0^t\frac{dM_{j}(s)}{E\{ Y_{j}(s)\}}, \ \frac{\left\{I(A=j) - \pi_j\right\} \Gamma_j(t)^\top  \Sigma^{-1}(W-\mux)}{\pi_j} \bigg] \\
& = {\rm Var} \left\{  S_j(t)\int_0^t\frac{dM_{j}(s)}{E\{ Y_{j}(s)\}}  \right\} + \frac{ (1- \pi_{j} )\Gamma_j(t)^\top \Sigma^{-1}  \Gamma_j(t) }{\pi_{j} }\\ 
& \quad - \frac{2 \Gamma_j(t)^\top \Sigma^{-1}}{\pi_{j} } S_j(t)E \left\{W \int_0^t\frac{dM_{j}(s)}{E\{ Y_{j}(s)\}} \right\}+ 2 \Gamma_j(t)^\top \Sigma^{-1} S_j(t)E \left\{W \int_0^t\frac{dM_{j}(s)}{E\{ Y_{j}(s)\}} \right\}\\
& = {\rm Var} \left\{  S_j(t)\int_0^t\frac{dM_{j}(s)}{E\{ Y_{j}(s)\}}  \right\} - \frac{ (1- \pi_{j} )\Gamma_j(t)^\top \Sigma^{-1}  \Gamma_j(t) }{\pi_{j}}.
\end{align*}

\section{Proof of Theorem \ref{prop:equiv}}

\paragraph{Equivalence between (a) and (b).} Let $\gamma_j=\Sigma^{-1}\mathbb{C}_j^g$. Define a bridge estimator 
\[
\tilde\theta_j = \frac{1}{n_j}\sum_{i:A_i=j}g(Y_{ij})-\gamma_j^\top(\bar W_j-\bar W).
\]
It is sufficient to show $\tilde\theta_j-\theta_j=n^{-1}\sum\psi_{ij}^g+o_p(n^{-1/2})$, $\hat \theta_j^{\rm \ AUG}-\tilde\theta_j=o_p(n^{-1/2})$, and $\hat \theta_j-\tilde\theta_j=o_p(n^{-1/2})$. It is straightforward that 
\begin{align*}
    \tilde\theta_j-\theta_j&=\frac{1}{n_j}\sum_{i:A_i=j}\{g(Y_{ij})-\theta_j\}-\gamma_j^\top(\bar W_j-\bar W)\\
    &=\frac{1}{\pi_jn}\sum_{i=1}^{n}I(A_i=j)\{g(Y_{ij})-\theta_j\}-\frac{1}{\pi_j}\gamma_j^\top\bar U_j+o_p(n^{-1/2})\\
    &=\frac{1}{n}\sum_{i=1}^n\frac{I(A_i=j)\{g(Y_{ij})-\theta_j\}-\{I(A_i=j)-\pi_j\}\gamma_j^\top(W_i-\mux)}{\pi_j}+o_p(n^{-1/2})\\
    &=\frac{1}{n}\sum_{i=1}^n\psi_{ij}^g+o_p(n^{-1/2}),
\end{align*}

\noindent where the second equation follows from $\bar W_j-\bar W=\pi_j^{-1}\bar U_j+o_p(n^{-1/2})$.

By the law of large numbers under (D), we have $\hat\Sigma_j=\Sigma+o_p(1)$ and $n_j^{-1}\sum_{i:A_i=j}(W_i-\bar W_j)g(Y_{ij})=\mathbb C_j^g+o_p(1)$. Therefore, $\hat\gamma_j=\gamma_j+o_p(1)$, which implies $\hat\theta_j^{\rm AUG}-\tilde\theta_j=-(\hat\gamma_j-\gamma_j)^\top(\bar W_j-\bar W)=o_p(n^{-1/2})$ using $\bar W_j-\bar W=O_p(n^{-1/2})$. 

It remains to show $\hat \theta_j-\tilde\theta_j=o_p(n^{-1/2})$. By lemma \ref{lemma1}, we have $n^{-1}\hat\lambda_j=\Sigma^{-1}\bar U_j+o_p(n^{-1/2})$. Using $\bar W_j-\bar W=\pi_j^{-1}\bar U_j+o_p(n^{-1/2})$ again gives $n_j^{-1}\hat\lambda_j=\Sigma^{-1}(\bar W_j-\bar W)+o_p(n^{-1/2})$. Recall our notation $s_{ij}=\hat\lambda_j^\top(W_i-\muxh)/n_j$ in the proof of lemma \ref{lemma1} and the result $\max_{i:A_i=j}|s_{ij}|=o_p(1).$ Then we expand the $\hat\theta_j$ as follows, with probability approaching 1, 
\begin{align*}
    \hat\theta_j&=\sum_{i:A_i=j}\hat p_{ij}g(Y_{ij})\\
    &=\frac{1}{n_j}\sum_{i:A_i=j}\frac{g(Y_{ij})}{1+s_{ij}}\\
    &=\frac{1}{n_j}\sum_{i:A_i=j}g(Y_{ij})\left\{1-s_{ij}+O(s_{ij}^2)\right\}\\
    &=\frac{1}{n_j}\sum_{i:A_i=j}g(Y_{ij})-\left(\frac{\hat\lambda_j^\top}{n_j}\right)\left\{\frac{1}{n_j}\sum_{i:A_i=j}(W_i-\muxh)g(Y_{ij})\right\}+R_{nj},
\end{align*}

\noindent where $R_{nj}=n_j^{-1}\sum_{i:A_i=j}g(Y_{ij})O(s_{ij}^2)$ and the third equation comes from $(1+s)^{-1}=1-s+O(s^2)$ when $s$ is in a small neighborhood around zero. For $R_{nj}$, we can see that 
\[
    |R_{nj}| \leq \max_{i:A_i=j}|s_{ij}|\frac{1}{n_j}\sum_{i:A_i=j}|g(Y_{ij})|\frac{\|\hat\lambda_j\|}{n_j}\|W_i-\muxh\|O(1)=o_p(1)O_p(n^{-1/2})O_p(1)=o_p(n^{-1/2}), 
\]
where we used $\hat\lambda_j/n_j=O_p(n^{-1/2})$ and $\frac{1}{n_j}\sum_{i:A_i=j}|g(Y_{ij})|\|W_i-\muxh\|=O_p(1)$ by Cauchy-Schwarz inequality. By the law of large number and Lemma \ref{lemma1}, we have $\frac{1}{n_j}\sum_{i:A_i=j}(W_i-\muxh)g(Y_{ij})=\mathbb C_j^g+o_p(1)$. Recall that $n_j^{-1}\hat\lambda_j=\Sigma^{-1}(\bar W_j-\bar W)+o_p(n^{-1/2})$. Together gives 
\[
\hat\theta_j=\frac{1}{n_j}\sum_{i:A_i=j}g(Y_{ij})-\mathbb C_j^{g\top}\Sigma^{-1}(\bar W_j-\bar W)+o_p(n^{-1/2})=\tilde\theta_j+o_p(n^{-1/2}), 
\]
which completes the proof of equivalence between (a) and (b).

\paragraph{Equivalence between (a) and (c).} It remains only to prove $\hat\theta_j^{\rm EB}=\tilde\theta_j+o_p(n^{-1/2}).$ Entropy balancing has a strictly convex finite-dimensional dual, and the corresponding primal weights have a normalized exponential form by the
KKT conditions \citep{zhao2017entropy}. Applying the same KKT calculation gives 
\[
\hat p_{ij}^{\rm EB}=\frac{\exp\{\hat\nu_j^\top(W_i-\bar W)\}}{\sum_{k:A_k=j}\exp\{\hat\nu_j^\top(W_k-\bar W)\}}, 
\]
where $\hat\nu_j$ is chosen so that $\sum_{i:A_i=j}\hat p_{ij}^{\rm EB}(W_i-\bar W)=0$. Equivalently, define 
\[
h_j(\nu)=\frac{n_j^{-1}\sum_{i:A_i=j}\exp\{\nu^\top(W_i-\bar W)\}(W_i-\bar W)}{n_j^{-1}\sum_{i:A_i=j}\exp\{\nu^\top(W_i-\bar W)\}} 
\]
with $h_j(\hat\nu_j)=0$. We expand $h_j(\nu)$ near $\nu=0$ under $\|\nu\|=O(n^{-1/2})$. Recall that \(\max_i\|W_i-\bar W\|=o_p(n^{1/2})\) and $\bar W_j-\bar W=O_p(n^{-1/2})$. We have \(\max_{i:A_i=j}|\nu^\top (W_i-\bar W)|=o_p(1)\). By using $\exp(x)=1+x+O(x^2)$ around $x=0$, the numerator of $h_j$ becomes 
\[
\frac{1}{n_j}\sum_{i:A_i=j}(W_i-\bar W)+\frac{1}{n_j}\sum_{i:A_i=j}(W_i-\bar W)(W_i-\bar W)^\top\nu+o_p(n^{-1/2}), 
\]
where the remainder is from \(O_p(\|\nu\|^2 n_j^{-1}\sum_{i:A_i=j}\|W_i-\bar W\|^3)=O_p(n^{-1})=o_p(n^{-1/2})\).
The first term is $\bar W_j-\bar W$ and the second is $\hat \Sigma_{j}\nu+(\bar W_j-\bar W)(\bar W_j-\bar W)^\top\nu=\hat \Sigma_{j}\nu+o_p(n^{-1/2})$, where $\hat\Sigma_j=n_{j}^{-1}\sum_{i:A_i=j}(W_i-\bar W_j)(W_i-\bar W_j)^\top$. So the numerator is \(\bar W_j-\bar W+\hat\Sigma_j\nu+o_p(n^{-1/2}).\) Similarly we can show the denominator of $h_j$ is \(1+O_p(n^{-1}).\) Therefore, 
\[
h_j(\nu)=\bar W_j-\bar W+\hat\Sigma_j\nu + o_p(n^{-1/2}). 
\]
To justify that the solution lies in this local region, let
\(Q_j(\nu)=\log\{n_j^{-1}\sum_{i:A_i=j}\exp[\nu^\top(W_i-\bar W)]\}\), so that \(\nabla Q_j(\nu)=h_j(\nu)\). Since \(\hat\Sigma_j\to_p\Sigma\) and \(\Sigma\) is positive definite, with probability tending to one there is a constant \(c>0\) such that the smallest eigenvalue of \(\hat\Sigma_j\) is larger than $c$. Also \(\|\bar W_j-\bar W\|=O_p(n^{-1/2})\). Hence, for \(\|\nu\|=Mn^{-1/2}\), 
\begin{align*}
\nu^\top h_j(\nu)&=\nu^\top(\bar W_j-\bar W)+\nu^\top\hat\Sigma_j\nu+\nu^\top o_p(n^{-1/2})\\
&\ge-\|\nu\|\,\|\bar W_j-\bar W\|+c\|\nu\|^2+\|\nu\|o_p(n^{-1/2})\\
&\ge n^{-1}\{cM^2-O_p(M)+o_p(M)\}.
\end{align*}

\noindent Choosing \(M\) large enough gives \(\nu^\top h_j(\nu)>0\) on the boundary
with probability tending to one. Since \(Q_j\) is convex and the EB
solution is unique, the minimizer lies inside this ball. Hence
\(\|\hat\nu_j\|=O_p(n^{-1/2})\).
Plugging into the expansion gives $0=h_j(\hat\nu_j)=\bar W_j-\bar W+\hat\Sigma_j\hat\nu_j+o_p(n^{-1/2})$ and therefore $\hat\nu_j=-\hat\Sigma_j^{-1}(\bar W_j-\bar W)+o_p(n^{-1/2})=-\Sigma^{-1}(\bar W_j-\bar W)+o_p(n^{-1/2}).$

Following how we expand $h_j$ and $\hat\theta_j$, we have 
\begin{align*}
\hat\theta_j^{\rm EB}&=\frac{1}{n_j}\sum_{i:A_i=j}g(Y_{ij})+\hat\nu_j^\top\left\{\frac{1}{n_j}\sum_{i:A_i=j}(W_i-\bar W_j)g(Y_{ij})\right\} + o_p(n^{-1/2})\\
&=\frac{1}{n_j}\sum_{i:A_i=j}g(Y_{ij})-\mathbb C_j^{g\top}\Sigma^{-1}(\bar W_j-\bar W) + o_p(n^{-1/2})\\
&=\tilde \theta_j+o_p(n^{-1/2}),
\end{align*}

\noindent which completes the proof.

\section{Proof of Theorem 3}
 The proof of Theorem 3(i)  follows from Theorem 1(i) and Bahadur's representation, e.g., Theorem 5.11 in \cite{Shao2003}. { Let $q_l = F^{-1}_l(p)$ and $\hat q_l = \hat F^{-1}_l(p)$. Then ${n}^{1/2} \{ \hat q_{j}- \hat q_k - q_j+ q_k\} = n^{-1/2}\sum_{i=1}^n \{-f_{jp}^{-1}\phi_{ij}(q_j)+f_{kp}^{-1}\phi_{ik}(q_k)\}+o_p(1)$ and the variance is
 \begin{align*}
      &\quad \frac{V_{F_j}(q_j)}{f_{jp}^2}+ \frac{V_{F_k}(q_k)}{f_{kp}^2}-\frac{2}{f_{jp}f_{kp}}{\rm Cov}(\phi_{ij}(q_j),\phi_{ik}(q_k))\\
      &=\frac{p(1-p)-(1-\pi_j)\C_j^\top\Sigma^{-1}\C_j}{\pi_jf_{jp}^2}+\frac{p(1-p)-(1-\pi_k)\C_k^\top\Sigma^{-1}\C_k}{\pi_kf_{kp}^2}-\frac{2\C_j^\top\Sigma^{-1}\C_k}{f_{jp}f_{kp}}\\
      &=\frac{p(1-p)}{\pi_jf_{jp}^2}+\frac{p(1-p)}{\pi_kf_{kp}^2}-\xi_{jk},
 \end{align*}
 where, with $\tilde\C_l=\C_l/f_{lp}$,
 \begin{align*}
\xi_{jk}&=\frac{1-\pi_j}{\pi_j}\tilde\C_j^\top\Sigma^{-1}\tilde\C_j+\frac{1-\pi_k}{\pi_k}\tilde\C_k^\top\Sigma^{-1}\tilde\C_k+2\tilde \C_j^\top\Sigma^{-1}\tilde\C_k\\
&=\left\{\frac{\pi_k}{\pi_j(\pi_j+\pi_k)}+\frac{1-\pi_j-\pi_k}{(\pi_j+\pi_k)}\right\}\tilde\C_j^\top\Sigma^{-1}\tilde\C_j+\left\{\frac{\pi_j}{\pi_k(\pi_j+\pi_k)}+\frac{1-\pi_j-\pi_k}{(\pi_j+\pi_k)}\right\}\tilde\C_k^\top\Sigma^{-1}\tilde\C_k\\
&\quad +(\frac{2}{\pi_j+\pi_k}-2\frac{1-\pi_j-\pi_k}{\pi_j+\pi_k})\tilde\C_j^\top\Sigma^{-1}\tilde \C_k\\
&=\frac{(\pi_k\tilde\C_j+\pi_j\tilde\C_k)^\top\Sigma^{-1}(\pi_k\tilde\C_j+\pi_j\tilde\C_k)}{\pi_j\pi_k(\pi_j+\pi_k)}+\frac{1-\pi_j-\pi_k}{\pi_j+\pi_k}(\tilde\C_j-\tilde\C_k)^\top\Sigma^{-1}(\tilde\C_j-\tilde\C_k).
 \end{align*}}
 The proof of Theorem 3(iii) follows from Theorem 1(ii) and the argument in \cite{sun2025}.
 For Theorem 3(ii), by Lemma 1, 
 \begin{align*}
 	\sqrt{n} (\hat \theta_{jk} - \theta_{jk})
 	% & = \sqrt{n} \{ \bar D_{jk} + \bar D_{kj} +  (\bar{\bX}_{j}-\bar{\bX})^\top \hat \bbeta_j 	- (\bar{\bX}_{k}-\bar{\bX})^\top \hat \bbeta_k \} 	+ o_p(1) \\ 
 	& = \sqrt{n} \{ \bar D_{jk} + \bar D_{kj}
 	+  (\bar{W}_{j}-\bar{W})^\top  \bbeta_j
 	- (\bar{W}_{k}-\bar{W})^\top  \bbeta_k \}
 	+ o_p(1) , 
 \end{align*}
 where
 $$ 
 \bar D_{jk} =  \frac{1}{n_j} \sum_{i:A_i=j} \{ 1- F_k(Y_{ij}) - \theta_{jk} \} , \qquad \bar D_{kj} = \frac{1}{n_k} \sum_{i:A_i=k} \{ F_j(Y_{ik}) - \theta_{jk} \} ,
 $$
 and $\beta_j = \Sigma^{-1} \mathbb{C}_{jk}$, $\beta_k = \Sigma^{-1} \mathbb{C}_{kj}$. 
 The rest of proof follows the same argument in the proofs of Corollary 1 and Theorem 3 in \cite{ye2021better}, since $\bar D_{jk}$ and $\bar D_{kj}$ are types of sample means with outcomes in treatment groups $j$ and $k$, respectively, and 
 \[  \zeta_{jk}= \lim_{n \to \infty} n\left\{  \frac{n (n_j\bbeta_k+ n_k\bbeta_j)^\top \bSigma 
 	(n_j\bbeta_k+ n_k\bbeta_j)}{ n_jn_k (n_j+ n_k) } 
 +  \frac{(n-n_j-n_k) (\bbeta_j-\bbeta_k)^\top \bSigma (\bbeta_j-\bbeta_k)}{(n_j+n_k)n} \right\} . \]

\section{Simulation results for censored outcome under  scenario 3 in \cite{zhang2015robust}}
 
 We provide simulation results under Section 6 with simulation setting changed to that in 
 scenario 3 in \cite{zhang2015robust}, where  
 covariate $X = (B, X_1,X_2,X_3)^\top$, $B$ is binary with $P(B=1)=0.5$, 
 $X_1$, $X_2$, and $X_3$ standard normal variables that are independent except that $X_3$ is correlated with $X_2$ with a correlation coefficient 0.3,  
 the  life time $T_j \sim \log N( 0.5 B + 0.5 X_2 + 0.6 X_3 + 0.3 (j-1), 1)$ (conditioned on $X$), 
 and the censoring time $C_j \sim $ uniform(0, 10)$+ 0.75 (j-1)$,  independent of $X$ and $T_j$, $j=1,2$. %Note $X_1$ does not predict survival.
 For assigning treatments,  we employ simple randomization and the stratified permuted block of size 4, where the strata are formed by $Z$ with two components, $B$ and $X_3$ discretized into four categories with equal probabilities based on the quartiles.
 The adjusted covariate used in estimation is $W= (B, X_{1c},X_{2c},X_{3c})^\top$ under simple randomization and $W= (Z, X_{1c},X_{2c},X_{3c})^\top$ under stratified permuted block randomization with $X_{lc}$ being $X_l$ truncated at $-5$ and 5. 
 
 The simulation results similar to those in Table1 of Section 6 are given in the following table, based on 2,000 replications. \vspace{5mm}

 \begin{table}[htbp!]
	\footnotesize
	\centering
	\begin{tabular}{cccccccccc}
		\hline
		&  &  &&& \multicolumn{5}{c}{Performance of estimation} \\
		Randomization & $j$ &  $t$&  $S_j(t)$ & \multicolumn{1}{c}{Estimator} & AB & SD & SE  & CP  & RE \\ \hline
		Simple & 1 & 1 & 0.573 & Kaplan-Meier & 0.001 & 0.050 & 0.051 & 0.949 &  \\
		 &  &  &  & Proposed MELE & 0.001 & 0.047 & 0.046 & 0.936 & 1.13 \\
		 &  &  &  & Zhang's AGEE & 0.001 & 0.047 & 0.046 & 0.940 & 1.14 \\
		 &  & 2 & 0.372 & Kaplan-Meier & 0.001 & 0.052 & 0.051 & 0.947 &  \\
		 &  &  &  & Proposed MELE & 0.001 & 0.049 & 0.046 & 0.933 & 1.13 \\
		 &  &  &  & Zhang's AGEE & 0.001 & 0.049 & 0.046 & 0.935 & 1.13 \\
		 &  & 3 & 0.267 & Kaplan-Meier & 0.000 & 0.049 & 0.048 & 0.953 &  \\
		 &  &  &  & Proposed MELE & 0.000 & 0.046 & 0.044 & 0.944 & 1.10 \\
		 &  &  &  & Zhang's AGEE & 0.000 & 0.046 & 0.044 & 0.945 & 1.10 \\
		 & 2 & 1 & 0.657 & Kaplan-Meier & 0.000 & 0.047 & 0.047 & 0.944 &  \\
		 &  &  &  & Proposed MELE & 0.001 & 0.043 & 0.043 & 0.940 & 1.15 \\
		 &  &  &  & Zhang's AGEE & 0.001 & 0.044 & 0.043 & 0.938 & 1.15 \\
		 &  & 2 & 0.458 & Kaplan-Meier & 0.001 & 0.051 & 0.051 & 0.946 &  \\
		 &  &  &  & Proposed MELE & 0.002 & 0.048 & 0.046 & 0.936 & 1.14 \\
		 &  &  &  & Zhang's AGEE & 0.002 & 0.048 & 0.046 & 0.937 & 1.15 \\
		 &  & 3 & 0.343 & Kaplan-Meier & 0.002 & 0.051 & 0.050 & 0.942 &  \\
		 &  &  &  & Proposed MELE & 0.002 & 0.049 & 0.045 & 0.929 & 1.10 \\
		 &  &  &  & Zhang's AGEE & 0.003 & 0.048 & 0.045 & 0.934 & 1.10 \\
		\cline{5-10}
		&&&& Zhang's AGEE & \multicolumn{5}{l}{0.05\% non-monotone estimated curves} \\
		\hline
		Stratified & 1 & 1 & 0.573 & Kaplan-Meier & 0.001 & 0.047 & 0.050 & 0.965 &  \\
		permuted &  &  &  & Proposed MELE & 0.001 & 0.046 & 0.045 & 0.946 &  \\
		block &  &  &  & Zhang's AGEE & 0.001 & 0.046 & 0.045 & 0.945 &  \\
		 &  & 2 & 0.372 & Kaplan-Meier & 0.001 & 0.050 & 0.051 & 0.955 &  \\
		 &  &  &  & Proposed MELE & 0.001 & 0.049 & 0.045 & 0.931 &  \\
		 &  &  &  & Zhang's AGEE & 0.001 & 0.049 & 0.045 & 0.930 &  \\
		 &  & 3 & 0.267 & Kaplan-Meier & 0.000 & 0.046 & 0.048 & 0.960 &  \\
		 &  &  &  & Proposed MELE & 0.001 & 0.046 & 0.043 & 0.933 &  \\
		 &  &  &  & Zhang's AGEE & 0.000 & 0.046 & 0.043 & 0.935 &  \\
		 & 2 & 1 & 0.657 & Kaplan-Meier & 0.000 & 0.046 & 0.048 & 0.962 &  \\
		 &  &  &  & Proposed MELE & 0.000 & 0.045 & 0.043 & 0.930 &  \\
		 &  &  &  & Zhang's AGEE & 0.001 & 0.045 & 0.043 & 0.931 &  \\
		 &  & 2 & 0.458 & Kaplan-Meier & 0.000 & 0.049 & 0.051 & 0.955 &  \\
		 &  &  &  & Proposed MELE & 0.000 & 0.048 & 0.046 & 0.933 &  \\
		 &  &  &  & Zhang's AGEE & 0.000 & 0.048 & 0.046 & 0.936 &  \\
		 &  & 3 & 0.343 & Kaplan-Meier & 0.001 & 0.049 & 0.050 & 0.956 &  \\
		 &  &  &  & Proposed MELE & 0.001 & 0.047 & 0.045 & 0.932 &  \\
		 &  &  &  & Zhang's AGEE & 0.001 & 0.047 & 0.045 & 0.933 &  \\
		\cline{5-10}
		&&&& Zhang's AGEE & \multicolumn{5}{l}{0.35\% non-monotone estimated curves} \\
		\hline
	\end{tabular}
\end{table}

\section{Simulation results for non-censored outcome} 

 We consider a simulation with $J=2$ to evaluate and compare the finite sample performances of  covariate adjusted MELE  $\hat F_1$ in (4) with $j=1$ and a univariate $X$, the unadjusted $\tilde F_1$ in (3), and the  covariate adjustment by regression   
	$ \hat F_1^{\rm A} $ in (5) \citep{rao1990}. 
	The simulation setting is simple randomization with $n_1 = n_2 = 100$,
	$X$ distributed as the chi-square distribution with one degree of freedom, and  $Y_1 $ distributed as $N(\beta X , \, 1)$ with $\beta =3$ conditioned on $X$.  
	
%	Based on 2,000 simulation replications, Table 1 gives the following results for 	estimating $F_1(y)$ at five different $y$'s,  the 10\%, 25\%, 50\%, 75\%, and 90\%  quantiles of $F_1$,  estimating the five quantiles, and  estimating the mean of $F_1$:  	the average of bias (AB),  standard deviation (SD), average of estimated SD (SE), coverage probability of 95\% asymptotic confidence interval (CP),  and relative efficiency (RE) defined as  (SD$^2$  of unadjusted)/SD$^2$.

We examine the estimation of $F_1(y)$ at $y =$  the  25\%, 50\%, and 75\% quantiles of $F_1$, the estimation of quantiles $F_1^{-1}(0.25)$, $F_1^{-1}(0.5)$, and $F_1^{-1}(0.75)$,  and the estimation of  mean of $F_1$. 
Based on 2,000 simulation replications, the following table gives 
the average of bias (AB),  standard deviation (SD), average of estimated SD (SE), coverage probability of 95\% Wald  confidence interval (CP),  and relative efficiency (RE) defined as  (SD$^2$  of unadjusted)/SD$^2$.
For each estimator, the SE is squared root of the variance estimator discussed in Section 5. For quantile estimators, the bootstrap described in \S1 is applied with $B= 500$.

The simulation findings are similar to those in Section 6. For quantile estimators, 
the bootstrap variance estimator work well. In every of 2,000 simulation runs, 
 the adjusted estimator $\hat F_1^{\rm A}$ defined by (5) is not monotone somewhere in the outcome range.

\begin{table}[H]
	\small
	\centering
	\begin{tabular}{cccccccc}
		\hline
		&  &  & \multicolumn{5}{c}{Performance of estimation} \\
		Estimand type & True parameter & \multicolumn{1}{c}{Estimator} & AB & SD & SE & CP & RE \\ \hline
		Distribution & $F_1(y_1)= 0.250$ & Unadjusted & 0.002 & 0.045 & 0.043 & 0.942 &  \\
		 &  & Proposed MELE & 0.001 & 0.044 & 0.042 & 0.930 & 1.08 \\
		 &  & Adjustment (5) & 0.001 & 0.044 & 0.042 & 0.932 & 1.04 \\
		\cline{3-8}
		 & $F_1(y_2)= 0.500$ & Unadjusted & 0.000 & 0.051 & 0.050 & 0.941 &  \\
		 &  & Proposed MELE & 0.000 & 0.046 & 0.045 & 0.945 & 1.19 \\
		 &  & Adjustment (5) & 0.001 & 0.044 & 0.045 & 0.954 & 1.30 \\
		\cline{3-8}
		 & $F_1(y_3)= 0.750$ & Unadjusted & 0.000 & 0.042 & 0.043 & 0.945 &  \\
		 &  & Proposed MELE & 0.000 & 0.036 & 0.036 & 0.944 & 1.39 \\
		 &  & Adjustment (5) & 0.000 & 0.035 & 0.036 & 0.955 & 1.49 \\
		\hline
		Quantile & $F_1^{-1}(0.25)= 0.363$ & Unadjusted & -0.023 & 0.216 & 0.219 & 0.933 &  \\
		 &  & Proposed MELE & -0.001 & 0.208 & 0.212 & 0.938 & 1.08 \\
		 &  & Adjustment (5) & -0.048 & 0.214 & 0.215 & 0.913 & 1.02 \\
		\cline{3-8}
		 & $F_1^{-1}(0.5)= 1.639$ & Unadjusted & -0.012 & 0.315 & 0.324 & 0.939 &  \\
		 &  & Proposed MELE & 0.010 & 0.288 & 0.296 & 0.940 & 1.20 \\
		 &  & Adjustment (5) & -0.024 & 0.271 & 0.275 & 0.931 & 1.36 \\
		\cline{3-8}
		 & $F_1^{-1}(0.75)= 4.120$ & Unadjusted & -0.065 & 0.689 & 0.737 & 0.932 &  \\
		 &  & Proposed MELE & 0.016 & 0.603 & 0.628 & 0.933 & 1.31 \\
		 &  & Adjustment (5) & -0.076 & 0.560 & 0.567 & 0.922 & 1.52 \\
		\hline
		Mean & $\int y\,dF_1(y)=3.000$ & Unadjusted & -0.009 & 0.433 & 0.427 & 0.933 &  \\
		 &  & Proposed MELE & -0.009 & 0.330 & 0.305 & 0.920 & 1.73 \\
		 &  & Adjustment (5) & -0.004 & 0.325 & 0.305 & 0.928 & 1.78 \\
		\hline
	\end{tabular}
\end{table}

\section{Values of estimates in the example of Section 7}

\begin{table}[hb]
\begin{center}
  {Table S1: Survival estimates and RMST difference at selected weeks}
\end{center}
\footnotesize
\centering
\begin{tabular}{lcccccc}
\hline
& & \multicolumn{2}{c}{Proposed MELE} & \multicolumn{2}{c}{Kaplan-Meier} & \\
& Week & estimate & SE & estimate & SE & SE$^2$ reduction \\ \hline

Control & 12 & 0.987 & 0.009 & 0.987 & 0.009 &  0.8\% \\ 
 & 24 & 0.973 & 0.013 & 0.974 & 0.013 &  1.9\% \\ 
 & 36 & 0.948 & 0.018 & 0.948 & 0.018 &  0.9\% \\ 
 & 48 & 0.874 & 0.027 & 0.875 & 0.027 &  1.3\% \\ 
 & 60 & 0.823 & 0.031 & 0.821 & 0.031 &  2.0\% \\ 
 & 72 & 0.812 & 0.032 & 0.808 & 0.032 &  1.9\% \\ 
 & 84 & 0.744 & 0.038 & 0.738 & 0.038 &  3.5\% \\ 
 & 96 & 0.672 & 0.048 & 0.667 & 0.048 &  2.4\% \\ 
\noalign{\medskip}
Treatment & 12 & 0.970 & 0.013 & 0.969 & 0.014 &  0.7\% \\ 
 & 24 & 0.948 & 0.017 & 0.950 & 0.017 &  1.3\% \\ 
 & 36 & 0.948 & 0.017 & 0.950 & 0.017 &  1.3\% \\ 
 & 48 & 0.933 & 0.019 & 0.937 & 0.019 &  2.0\% \\ 
 & 60 & 0.914 & 0.022 & 0.917 & 0.022 &  2.3\% \\ 
 & 72 & 0.898 & 0.023 & 0.903 & 0.024 &  2.4\% \\ 
 & 84 & 0.849 & 0.031 & 0.851 & 0.032 &  2.6\% \\ 
 & 96 & 0.800 & 0.042 & 0.818 & 0.043 &  4.2\% \\ 
\hline 
RMST diff & 100 & 5.348 & 2.563 & 5.837 & 2.640 &  5.7\% \\ 
\hline \noalign{\medskip}
\multicolumn{7}{l}{SE$^2$ reduction = (SE$^2$ of Kaplan-Meier $-$ SE$^2$)/(SE$^2$ of Kaplan-Meier)}
\end{tabular}
\end{table}

\bibliographystyle{apalike}
\bibliography{ms}

\end{document}